\documentclass[useAMS,fleqn,usenatbib]{mnras}

\usepackage{multirow}
\usepackage{url}
\usepackage{hyperref}
\usepackage{ulem}

\usepackage[T1]{fontenc}

\usepackage{ae,aecompl}
\usepackage{graphicx}	
\usepackage{amsmath}	
\usepackage{amssymb}	
\usepackage{threeparttable}
\usepackage{booktabs}

\usepackage{xcolor,listings}

\usepackage{wasysym} 
\usepackage{scalerel} 
\usepackage{verbatim} 

\definecolor{darkgreen}{rgb}{0.0,0.5,0.0}
\definecolor{darkred}{rgb}{0.5,0.0,0.0}
\definecolor{brown}{rgb}{0.8,0.4,0.0}

\def\apj {ApJ}
\def\apjl {ApJL}
\def\apjs {ApJS}
\def\aj {AJ}
\def\aap {A\&A}
\def\mnras {MNRAS}

\def\nat {Nature}

\def\Msol {{\cal M}_{\odot}}

\def\R200 {R_{200}}

\def\mpc{h^{-1} {\rm{Mpc}}}
\def\kms {\rm{km~s^{-1}}}
\def\Mpc {\rm Mpc}

\defcitealias{tempel+17}{T17}
\usepackage{newtxtext,newtxmath}

\begin{document}


\title[Compact groups inhabiting different environments]{Hickson-like compact groups inhabiting different environments}

\author[Taverna et al.]{\parbox[t]{\textwidth}{
A. Taverna\thanks{ataverna@unc.edu.ar}, J.M. Salerno, I.V. Daza-Perilla, E. D\'iaz-Gim\'enez, A. Zandivarez, H.J. Mart\'inez, A.N. Ruiz
}\\
 \\
1. CONICET. Instituto de Astronom\'ia Te\'orica y Experimental (IATE), Laprida 854, X5000BGR, C\'ordoba, Argentina\\
2. Universidad Nacional de C\'ordoba (UNC). Observatorio Astron\'omico de C\'ordoba (OAC), Laprida 854, X5000BGR, C\'ordoba, Argentina\\
}
\date{Accepted XXX. Received YYY; in original form ZZZ}

\pubyear{2023}


\label{firstpage}
\pagerange{\pageref{firstpage}--\pageref{lastpage}}
\maketitle
 
\begin{abstract}
Although Compact Groups of galaxies (CGs) have been envisioned as isolated extremely dense structures in the Universe, it is accepted today that many of them could be not as isolated as thought.
In this work, we study Hickson-like CGs identified in the Sloan Digital Sky Survey Data Release 16 to analyse these systems and their galaxies when embedded in different cosmological structures. To achieve this goal, we identify several cosmological structures where CGs can reside: Nodes of filaments, Loose Groups, Filaments and cosmic Voids. 
Our results indicate that 45 per cent of CGs do not reside in any of these structures, i.e., they can be considered non-embedded or isolated systems. 
Most of the embedded CGs are found inhabiting Loose Groups and Nodes, while there are almost no CGs residing well inside cosmic Voids. 
Some physical properties of CGs vary depending on the environment they inhabit. CGs in Nodes show the largest velocity dispersions, the brightest absolute magnitude of the first-ranked galaxy, and the smallest crossing times, while the opposite occurs in Non-Embedded CGs. 
When comparing galaxies in all the environments and galaxies in CGs, CGs show the highest fractions of red/early-type galaxy members in most of the absolute magnitudes ranges. The variation between galaxies in CGs inhabiting one or another environment is not as significant as the differences caused by belonging or not to a CG. 
Our results suggest a plausible scenario for galaxy evolution in CGs in which both, large-scale and local environments play essential roles.
\end{abstract}


\section{Introduction}

The large-scale structure of the Universe in the $\Lambda$ cold dark matter model ($\Lambda$CDM) is characterised by the anisotropic structure of the matter distribution. The matter instead tends to aggregate into complex structures that form a network called the cosmic web \citep*{Bond96}. As the Universe evolves, mass is accreted onto the densest concentrations forming even denser clumps, where galaxy clusters/groups are formed. These clumps give rise to regions almost devoid of galaxies named cosmic voids (\citealt*{einasto+80,van_de_voids+11}). Filaments trace the cosmic web and can be seen extending over scales up to tens of megaparsecs
(e.g., \citealt*{colberg+05}
). Even when voids represent most of the volume of the Universe, they contain only about $7$ per cent of the galaxies \citep{Pan+12}. Within this intricate cosmic network, galaxies are born, grow and evolve, each one following a particular evolutionary history that strongly depends on the environment in which its lifetime takes place.

It is well known that galaxy properties such as star formation, morphology, luminosity, colour, gas content and the structure of their subsystems correlate with the environment (e.g., \citealt*{dressler80,gomez03,martinez08}, \citealt{pandey+20,Bhambhani+22}). 
Pre-processing by galaxy groups/clusters and filaments has been extensively studied over the last decades both based on numerical simulations and observations. 
Galaxy clusters and massive groups are the places of several quenching processes that generate a population of quiescent galaxies (i.e., red/passive galaxy population). Most of these processes depend on the cluster mass. The external processes that shut down the star formation are called environmental quenching, while mass quenching refers to any internal process associated with the galaxy stellar mass (e.g., \citealt{Baldry04, Park07,Peng10}).
At intermediate density environments (filaments and low-mass loose groups), galaxies are bluer than cluster members, but yet are redder than their counterparts in the field and voids in the local Universe (\citealt{zhang13}, \citealt*{Martinez16}) and of those in intermediate redshifts ($0.4 < z < 0.9 $) \citep{Salerno2019, Sarron19}. These differences are most likely due to a combination of internal and external quenching processes.
On the other hand, low-density environments such as the field or voids, are dominated by blue/star-forming galaxies. 
As revealed by observational studies using void samples, galaxies in voids are bluer, have higher specific star formation rates and are of later types than galaxies living in regions at average density (e.g., \citealt{2004ApJ...617...50R, 2005ApJ...624..571R, 2006MNRAS.372.1710P} \citealt*{2008MNRAS.384.1189V, 2012MNRAS.426.3041H}, \citealt{rodriguez-medrano+22}). In both, field and voids, galaxies are likely to be quenched by internal mechanisms.

Among the different places in the Universe where galaxies inhabit, compact groups of galaxies (hereafter CGs) constitute a very interesting environment. 
They are dense galaxy systems containing a few luminous galaxies in close proximity to each other \citep*{hickson82} and relatively isolated from other bright galaxies. They are a high-density sub-category of poor groups and have typical sizes of a few tens of kiloparsecs in projection. Their densities are among the highest observed. In these small systems, galaxies are expected to strongly interact among themselves since their velocity dispersion is lower than those seen in massive loose groups or clusters of galaxies \citep{hickson92}. Several authors suggested that the CG environment accelerates the evolution of galaxies from star-forming to quiescent (e.g., \citealt{Tzanavaris10, Walker10}).

During the last forty years, several studies have been carried out using CGs as their main laboratory to study galaxy evolution. Among the latest studies, we can mention the work of \cite*{coenda12}. They performed a detailed comparison between the properties of galaxies in CGs and loose groups. These authors found that galaxies in CGs are, on average, systematically more concentrated, smaller in size, and have higher surface brightness than galaxies in the field or in loose groups. \cite*{Coenda15} also found that galaxies in CGs have an older population than loose groups or field galaxies. They concluded that CGs are extremely favourable environments for processes that transform star-formation galaxies into passive galaxies, and this transition is more efficient and faster than in other environments.
\cite{Lim17} studied mid-infrared (MIR) properties of galaxies inside CGs from the volume-limited catalogue of \citet{Sohn+16} and they compared these properties with those of galaxies in the field and clusters. 
They found that early-type galaxies in CGs are older than those of galaxy clusters, and the fraction of early-type depends on the environment. In addition, they concluded that CG environments play a critical role in accelerating morphology transformation and star formation quenching for the member galaxies, being this the best place for the pre-processing.
Recently, \cite*{zandivarez+22} analysed the influence of the Hickson-like CG environments on the luminosity of their galaxy members. They observed a brightening in the magnitudes of galaxies in CGs compared to galaxies in loose groups, and a deficiency of faint galaxies in CGs in comparison with loose groups. In addition, they showed that the luminosities of blue and late-type galaxies in CGs are equally affected as a function of the group virial mass as the luminosities of red and early-type galaxies, in contrast to what happens in loose groups where only the luminosities of red and early-type galaxies show a dependence with group virial mass. These authors suggested that the inner extreme environment in CGs may lead to different evolutionary histories for their galaxies.

Despite the numerous works about CGs, studies about their surrounding environment are less common, and even less frequent is the study of CGs inhabiting filaments and/or voids. Although CGs are meant to be isolated, the isolation is only relative to their own small sizes and regarding other bright galaxies. 
\cite{mendel+11} used a large sample of CGs and of galaxy groups to show that half of the CGs are associated with rich groups (or clusters). The other half were either independently distributed structures within the field (i.e. they are not embedded) or associated with relatively poor structures.

On the other hand, \cite*{diaz15} analysed samples of local CGs and loose groups.
They concluded that only 27 per cent  of the CGs can be considered embedded in larger galaxy systems.
\cite{Sohn+15} studied the local environment of CGs using the surface number density $\Sigma_5$. They assumed that the local environment of CGs is bimodal. Groups with high $\Sigma_5$ are considered embedded and CGs with low $\Sigma_5$ are isolated. They found that only $\sim  9$ per cent of CGs are embedded in a denser region. In a posterior work, using a Friends-of-Friends identifier of CGs, \cite{Sohn+16} defined the local environment by counting the number of neighbours of CGs. They found that group properties depend on the environment and they observed a larger fraction of early-type galaxies in dense environments with respect to low-density environments.

More recently, \citet{Zheng+21} 
classified CGs as embedded and not embedded in galaxy systems
and they found that half of their sample of CGs are embedded systems. Their results indicate that the dynamical properties of embedded CGs might depend on their parent groups, but, they could not distinguish if a CG is infalling into the group or if it is just a chance alignment.
\cite{Duplancic20} studied the impact of the environment on the galaxies of small systems. They used a catalogue of pairs, triplets and small galaxy groups (4 to 6 members
). These small systems were meant to have compactness similar to Hickson compact groups. They studied the environment of the system by counting the neighbours on different scales. They calculated the distance to the nearest filament to know the position in the cosmic web and found that their small galaxy groups are located in void walls and associated with long filaments, while their pairs and triplets are located in void environments.
With a very different approach, \cite{taverna+22} analysed Hickson-like compact groups extracted from semi-analytical models, and predicted that nearly 90 per cent  of the observational CGs are likely embedded in larger galaxy systems.

Although there are studies of CGs embedded in particular environments, a comprehensive study considering multiple environments at the same time is needed. 
Therefore, this paper aims to study the effects of different global environments on the properties of Hickson-like CGs and their galaxy members. We study the population of CGs that can be considered inhabiting galaxy groups, filaments and voids extracted from Sloan Digital Sky Server Data Release 16 (SDSS DR16, \citealt{DR16_sdss}). 
There are several methods to identify loose groups (i.e., \citealt{Huchra82,Merchan+02,LG_Yang2005,DuarteMamon+14,LG_facu2020}),  filaments (i.e., \citealt*{novikov+06}, \citealt{aragon-calvo+07,tempel+14_filaments,Pereyra19,buncher_carrasco+20,carron_duque+22}), and cosmic voids (i.e., \citealt{void_cecca+06}, \citealt*{platen+07,neyrinck+08}, \citealt{ruiz+15}). In this work, we follow the procedures of \cite{zandivarez+22}, \cite{Martinez16}, and \cite*{ruiz+19} to identify loose groups, filaments and voids, respectively.

This paper is organised as follows: in Section \ref{sec:data} we present all the samples of different structures used in this work. In Section \ref{sec:cg_env} we describe the procedures for associating CGs to each structure. In Section \ref{sec:results} we show the comparison among properties of CGs into these different structures. Finally, in Section \ref{sec:conclu} we summarise our results and present our conclusions. In this paper, we adopt a Planck cosmology \citep{Planck+16} with parameters: $h=0.67$ (dimensionless $z=0$ Hubble constant), $\Omega_{m}=0.315$ (matter density parameter), and $\sigma_8=0.83$ (standard deviation of the power spectrum on the scale of 8 $h^{-1}$ Mpc).

\section{Sample Selection}
\label{sec:data}

We use the sample of galaxies from SDSS DR16 spectroscopic catalogue (\citealt{DR16_sdss}) revisited by \cite{zandivarez+22}. They selected only those galaxies in the main contiguous area of the survey (the Legacy Survey), and extended the sample by adding galaxies with redshifts from other sources compiled by \cite{tempel+17} which is based on the SDSS DR12 catalogue \citep{DR12a,DR12b}. In addition, they completed the sample by adding galaxies whose redshifts were obtained from the NASA/IPAC Extragalactic Database (NED), and removing misclassified galaxies following the procedure described in \citealt*{DiazGimenez+18} and \citealt{zandivarez+22}.
The final flux-limited sample (hereafter DR16+) contains $565\,286$ galaxies with redshifts less than $0.2$, 
and model apparent magnitude less than $17.77$. Figure \ref{fig:v1} shows their rest-frame r-band absolute magnitudes as a function of redshifts. K-corrections were determined using the calculator developed by \cite{blanton_kcorr_07} at $z=0$.

From the flux-limited sample, we build a volume-limited sample of galaxies with $ z \le 0.1$ and $M_r \le -19.769$ (hereafter $V1$). 
In Fig.~\ref{fig:v1} we show the V1 sample delimited by dashed lines. These limits are adopted from the void identification process, which uses volume-limited samples to select structures (see section~\ref{sect:iden_voids} for more details).  

All the different structures used in this work are identified on the same parent catalogue DR16+. In the following subsections, we describe the main steps for the identification of all the involved cosmological structures: CGs, filaments and their nodes, loose groups, and voids.

\subsection{The sample of Compact Groups}
\label{sec:idencg}

We use the publicly available sample of CGs identified by \cite{zandivarez+22}\footnote{\url{https://cdsarc.cds.unistra.fr/viz-bin/cat/J/MNRAS/514/1231}}
on the DR16+ sample.
CGs were identified using Hickson-like criteria, i.e., a CG must satisfy the following criteria:

\begin{itemize}
\item {\bf Population}: $ \ 3 \le N \le 10$ 
\item {\bf Compactness}: $ \ \displaystyle \mu_r \le 26.33 \ {\rm [mag/arcsec^2]}$
\item {\bf Isolation}: $\ \displaystyle \Theta_N > 3 \,\Theta_G$
\item {\bf Flux limit}: $ \ \displaystyle r_{\rm bri} \le r_{\rm lim} - 3$
\item {\bf Velocity filtering}:   $ \ \displaystyle c  \frac{|z_i - z_{\rm cm}|}{1+z_{\rm cm}}  \le  \displaystyle  1000 \, \rm km \, s^{-1} $ 
\end{itemize}

\noindent where, $N$ is the number of galaxies within a three-magnitude range from the brightest galaxy in the $r$-band magnitude; $\mu_r$ is the mean $r$-band surface brightness averaged over the smallest circle that circumscribes the galaxy centres; $\Theta_{\rm G}$ is the angular diameter of the smallest circumscribed circle; $\Theta_{\rm N}$ is the angular diameter of the largest concentric circle that contains no other galaxies within the considered magnitude range or brighter; $r_{\rm bri}$ is the apparent magnitude of the brightest galaxy of the group; $r_{\rm lim}=17.77$ is the apparent magnitude limit of the parent catalogue; $c$ is the speed of light, $z_i$ is the spectroscopic redshift of the each galaxy members,  and $z_{\rm cm}$ is the bi-weighted median of the redshifts of the galaxy members.
We use the sample of CGs labelled as free from any potential source of contamination, i.e., $1412$ CGs with $4633$ galaxy members. 

\begin{figure}
\begin{center}
\includegraphics[width=\hsize]{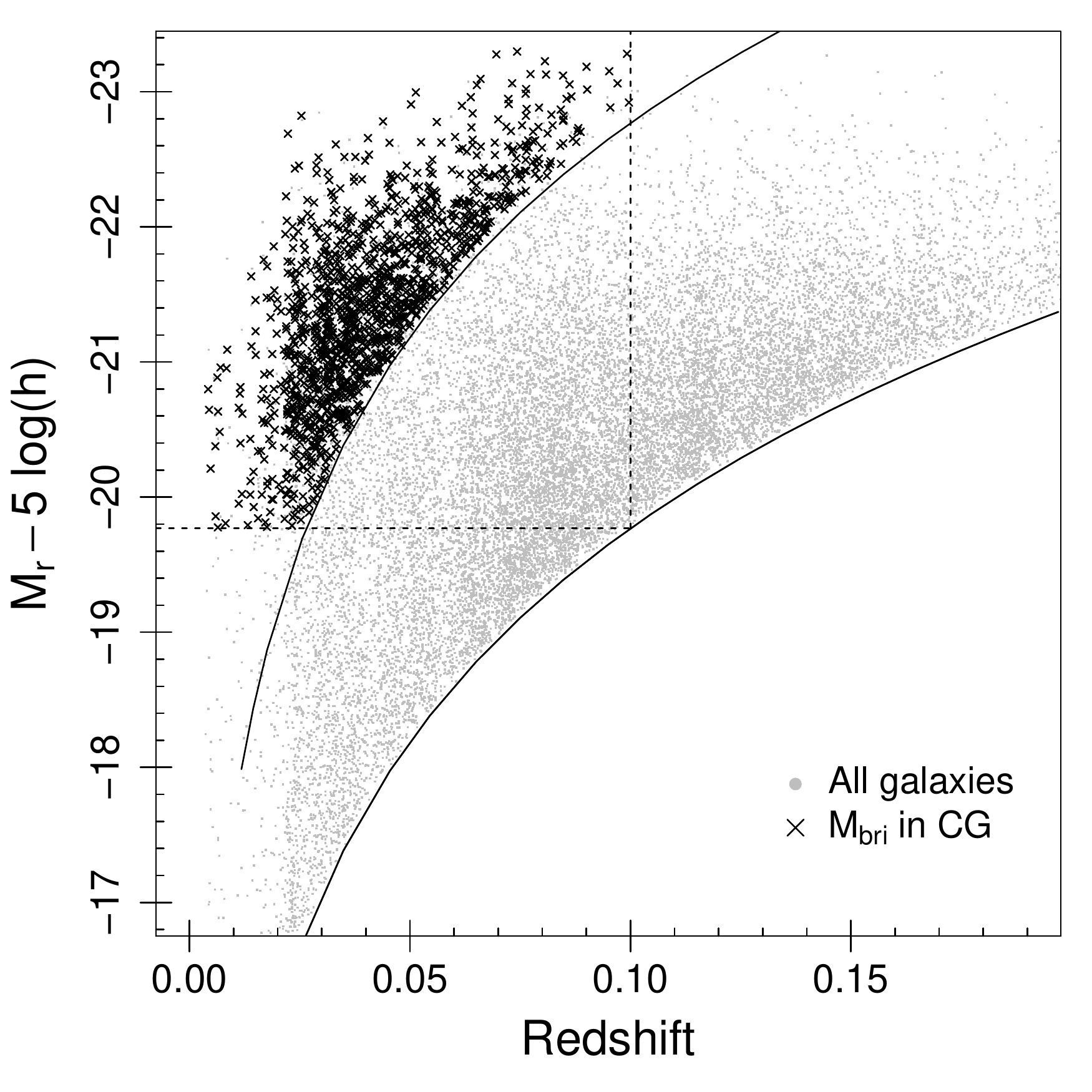}
\caption{r-band absolute magnitude versus redshift. Grey points are galaxies in the flux-limited sample (only 10 per cent  is shown). 
The lower solid line is computed from the r-band apparent magnitude limit of the parent catalogue DR16+, while the upper solid line is defined from the r-band apparent magnitude limit imposed tho the first-ranked galaxy in CGs (flux limit criterion).
Vertical and horizontal dashed lines enclose the volume-limited sample, V1. 
Dark crosses denote the first-ranked galaxies of CGs in V1.
}
\label{fig:v1}
\end{center}
\end{figure}

From this sample, we select CGs within the volume-limited sample, V1, to avoid introducing dependence on redshifts. 
To this end, we selected those CGs whose first-ranked galaxy absolute $r$-band magnitude is brighter than $-19.769$ and the group bi-weighted median redshift is $ z_{\rm cm} \le 0.1$. The final sample of CGs in V1 comprises $1368$ systems.
Since galaxy members are restricted to those galaxies within a range of three magnitudes from the brightest group galaxy, we complete the sample of galaxies in CGs by adding fainter galaxies that lie within the isolation cylinder around the group centre (i.e, angular distance less than $3\Theta_G$ and within $1000 \, \rm km/s$ from the group centre). Therefore, the sample of galaxies in CGs in V1 comprises $5551$ objects\footnote{We will refer to the original members of the CGs (without faints) as ``bright members''}.
In Fig.~\ref{fig:v1}, the first-ranked galaxies of CGs in V1 are superimposed (black crosses). Notice that the brightest galaxies of CGs are above an envelope (upper solid line) given by the CG flux limit criterion ($r_{\rm bri} \le r_{\rm lim} - 3$).

\subsection{The sample of Galaxy Groups}
\label{fof_sam}

We use the sample of galaxy groups identified by \cite{zandivarez+22} in the DR16+ flux-limited catalogue. 
To perform the identification, they applied a standard friends-of-friends algorithm (FoF, \citealt{Huchra82}) which uses two linking lengths to associate galaxies: one linking length in the line of sight ($V_L$), and another in projection in the sky ($D_L$).
The algorithm links pairs of galaxies that are separated less than the linking lengths in the projected distance and in the line-of-sight velocity. All galaxies which fulfil the conditions are associated by the algorithm and form a group.

The transverse and radial linking lengths, $D_L$ and $V_L$, are scaled with a factor to compensate for the different sampling of the luminosity function at different redshift \citep{Huchra82,Merchan+02,Eke04}. 
They identified loose galaxy groups using a transversal linking length $D_0 = 238 \, h^{-1} \rm kpc$ (which depends mainly on the Planck cosmological model) and a radial velocity threshold $V_0 =285 \, \rm km/s$ (which depends mainly of an analytical prescription for the redshift distortions) at a fiducial velocity of $1000\, \rm km/s$, and using the luminosity function of galaxies in the DR16+ determined in the same work to compute the scale factor\footnote{See subsection 2.3 of \cite{zandivarez+22} for a detailed description of the computation of the linking length parameters.}. The sample comprises $14\,652$ FoF galaxy groups. 

In addition, nothing prevents a CG be also identified as a FoF group by the search algorithm. However, a plain member-to-member comparison to detect FoF groups that are also CGs is not fair since the CG definition only considers as members those galaxies within a three-magnitude range from the brightest galaxy while the members of FoF groups are only limited by the apparent magnitude cut-off of the parent catalogue.
Hence, we adopt the following criteria to decide if a FoF group is already considered in the CG catalogue: 

\begin{enumerate}
    \item The CG shares more than $75$ per cent of its bright members with the FoF group.
    \item The number of relatively bright FoF galaxy members outside the CG isolation disk does not exceed half the number of bright members of the CG, 
    i.e, the number of FoF galaxies that lie in projection outside $3\Theta_G$ and whose magnitudes are brighter than $r_{\rm bri} + 3$ is at most half the number of bright members in the CG 
\end{enumerate}

\begin{figure}
\begin{center}
\includegraphics[width=0.69\columnwidth]{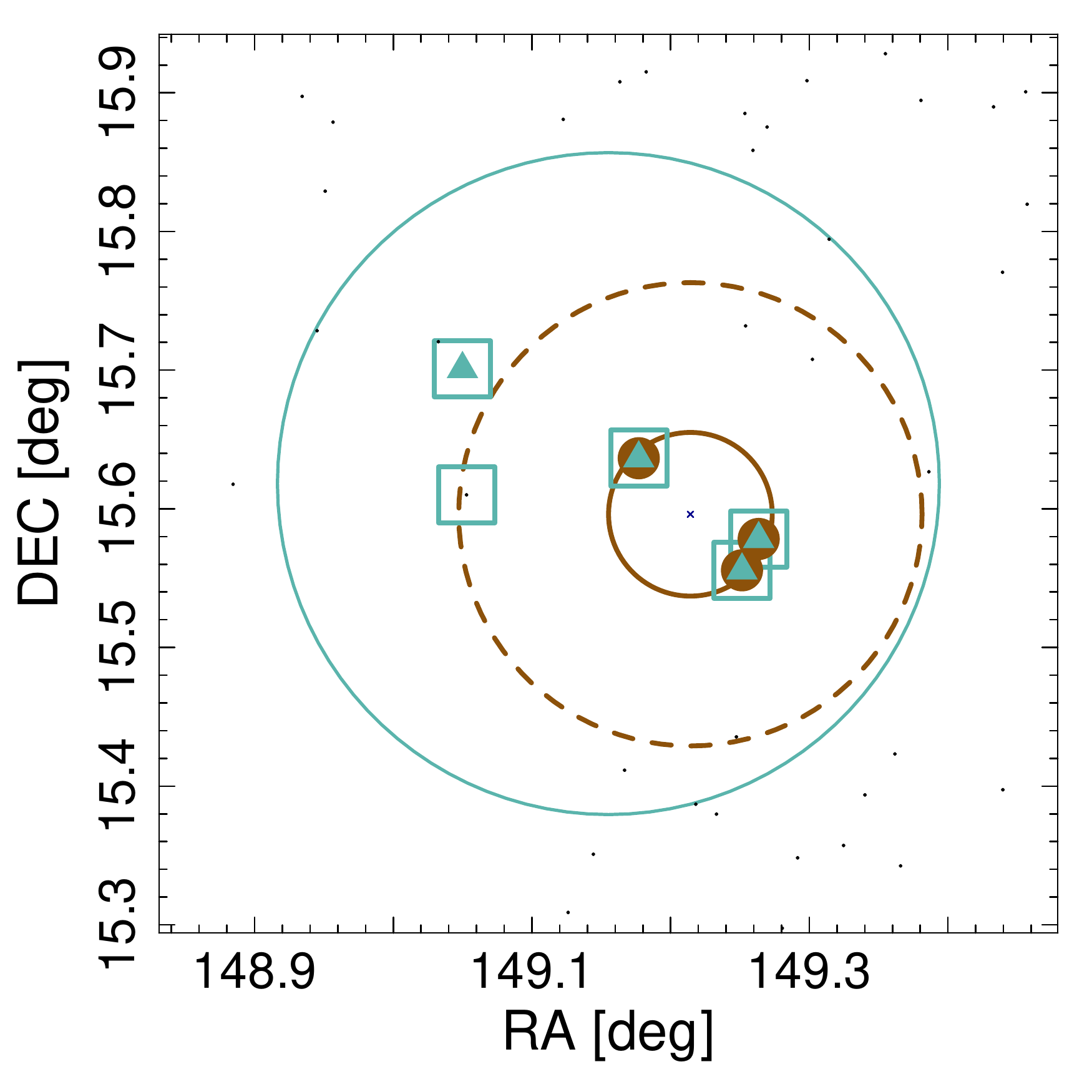}
\includegraphics[width=0.69\columnwidth]{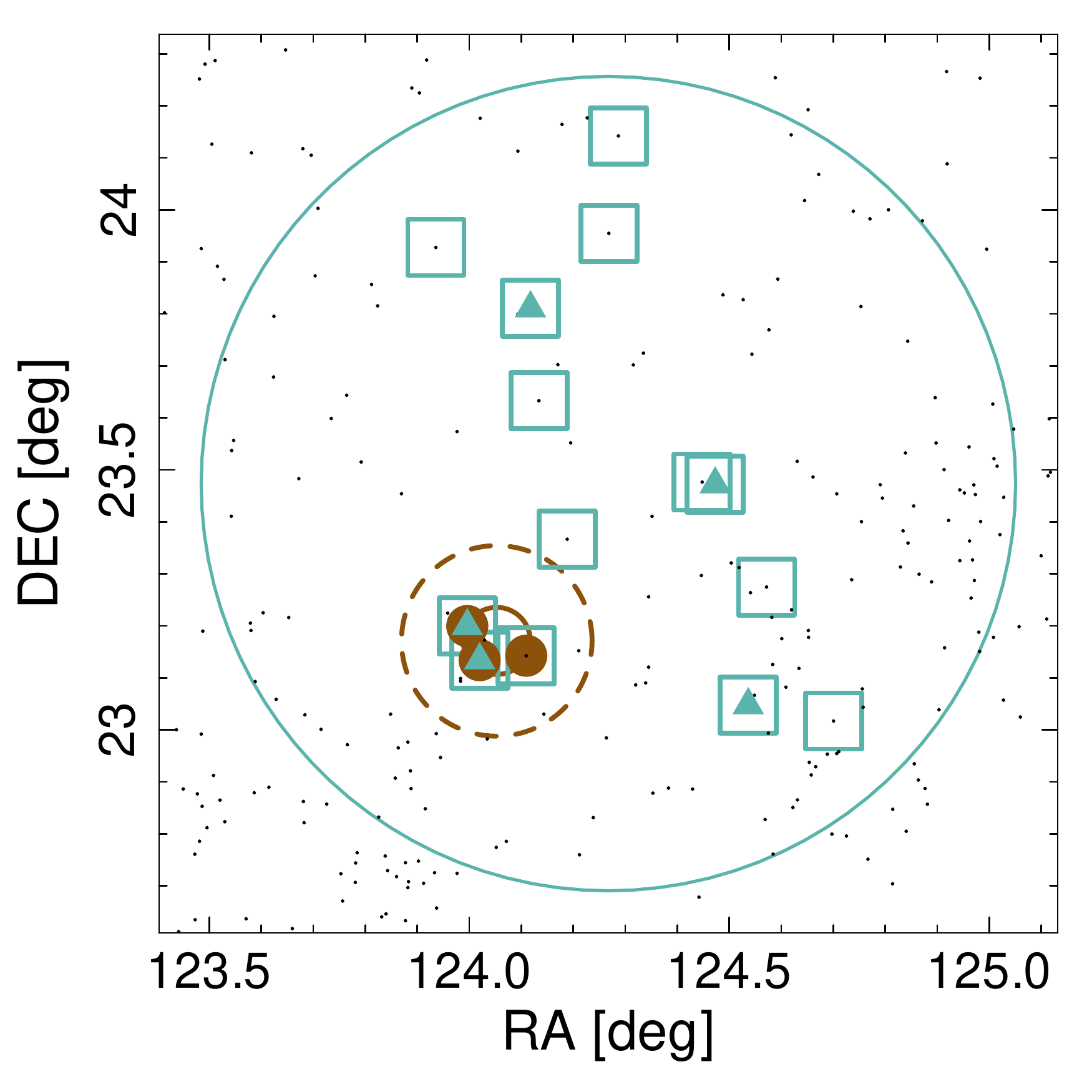}
\caption{Two examples of FoF groups in projection and their associated CGs. Field galaxies are represented as black dots; members of CGs are filled brown circles; FoF members are open cyan squares (filled cyan triangles represent galaxies within a three-magnitude range from the CG brightest galaxy). Brown solid circles represent the minimum circle that encloses all the CG members ($\Theta_G$), and, brown dashed circles indicate 3 times $\Theta_G$ (isolation area), while the cyan solid circles show the projected virial radii ($R_{\rm vir}$) of the FoF groups. {\it Top panel}: Example of a FoF group considered equal to a CG. All the members of the CG are also members of the FoF group. One of the remaining FoF members is a fainter galaxy, and only one bright galaxy is outside the CG isolation area. {\it Bottom panel}: Example of an FoF group hosting a CG. $75$ per cent of the members of the CG are also members of the FoF group, but the number of FoF bright members (triangles) outside the CG isolation area is larger than half the membership of the CG.
}
\label{fig:cg_fof}
\end{center}
\end{figure}

In those cases, we considered that the CG and the FoF group are the same system, and then we will not consider the system as a FoF group.
In the upper panel of Fig.~\ref{fig:cg_fof} we show an example of a FoF group and a CG considered the same system, while in the bottom panel we show an example of a FoF group hosting a CG, but they are not the same system. 
After applying these criteria, we remove from the sample of FoF groups $\sim 400$ systems that can be considered CGs. 
The clean sample of FoF groups comprises $14\,253$ systems
with at least 4 members and a median line-of-sight velocity dispersion, virial mass and 3D virial radius of $\langle \sigma_v\rangle=246 \, \kms$, $\langle {\cal M}_\textup{vir}\rangle =4.08 \times 10^{13} \, h^{-1} \Msol$, and $\langle R_\textup{vir}\rangle= 1.03 \, h^{-1} \Mpc $, respectively\footnote{The 3D velocity dispersion ($\sigma$) is estimated using the line-of-sight velocity dispersion $\sigma_v$,  $\sigma=  \sqrt{3}\, \sigma_v $, where $\sigma_v$ is computed with the biweight ($N\ge 15$) or gapper ($N<15$) scale estimators \citep*{Beers90}, and the 3D virial radius is computed as $R_{vir}=\frac{\pi}{2} R_{p}$, where the projected virial radius $R_{p}=2 \ \langle d_{ij}^{-1} \rangle^{-1}$ is twice the harmonic mean projected separation. The group virial masses are computed as ${\cal M}_{\rm vir}=\sigma^2 R_{\rm vir}/G$.}.

\subsubsection{The sample of Nodes and Filaments}
\label{sect:fil_sam}
We identify filamentary structures that extend between groups of galaxies following a procedure
based on that of \citet{Martinez16}.
The Filament identification is performed on the DR16+ sample.
The method starts with galaxy group pairs ($i,j$) as candidates to be Filament Nodes. We consider that a group pair is linked by a Filament when: 
\begin{enumerate}
    \item the comoving distance (2D+1/2) between the group centres is smaller than a given threshold $\Delta_{\textup{max}}$ and larger than the sum of their projected virial radii;
    \item the galaxy number density in a cylinder defined in redshift space with 
    the groups $i$ and $j$ at its ends doubles the mean galaxy number density at the pair’s redshift. These cylinders have a fixed radius $H$;
    \item the number of galaxies in the filamentary region, (i.e. the cylinder) is at least ten. 
\end{enumerate}

The main difference with \citet{Martinez16} is to
consider that Filaments linking groups are straight in shape and contained within 
cylindrical volumes stretching between the groups involved. In contrast to that paper, 
we intend to identify which galaxies are in Filaments, therefore our 
choice is much more restrictive. 
\citet{Martinez16} used larger, cuboid-like volumes between
groups which contained not only filament galaxies but also interlopers. 
\citet{Rost20} successfully used cylinders linking the original sample of nodes of \citet{Martinez16} to study a number of properties of Filaments.
In this work, we restrict the sample of Node candidates to massive and rich galaxy groups. 
With this restriction, we intend to identify Filaments that are much more likely to 
be real overdensities of galaxies stretching between systems. 

From the galaxy groups identified in the previous subsection, we select those systems with virial masses above the median mass of the sample 
(i.e., ${\cal M}_\textup{vir} \ge 10^{13.6} \, \Msol \, h^{-1}$) and population of at least ten galaxies.
We choose $\Delta_{\mathrm{max}}=14 \mpc$, which is slightly greater than
the redshift space correlation length for groups in this mass range according to 
\citet*{Zandivarez:2003}. The cylinder radius is chosen to be fixed and takes the 
value $H=1.5 \mpc$, which is the scale length that encloses 99 per cent  of all
galaxies in groups in projection.

The overdensity is computed by counting the number of galaxies
and points from a random catalogue that lay within the cylinder. 
Our random catalogue is approximately 100 times denser than its observational counterpart
and is constructed out of the latter using the method proposed by \citet{Cole:2011}\footnote{By construction, the \citet{Cole:2011} method produces a random catalogue that is very close to be
as denser as desired but not exactly.} 
for assigning
redshifts to the random points, and an angular mask for the footprint of the DR16+\footnote{The angular mask for the DR16+ has been derived using the HEALPix \citep{healpix} package (\url{ http://healpix.sourceforge.net}).}.
To exclude group galaxy members when counting galaxies in the cylinder, 
we remove from the computation all objects within 
a projected radius of $r_p\leq 1.1 \times R_p$ and line-of-sight velocity difference $|\Delta v|\leq 
3\times \sigma_v$ from each group's centre. This excludes $>99$ per cent of all group galaxies. 
Out of consistency, we do the same over the random sample.
The overdensity is defined as $\Delta_f = (N_r n_{g}/N_g n_{r}) - 1$, where $n_{g}$ ($N_g$)
and $n_{r}$ ($N_r$) are the number of galaxies and of random points within the cylinder (in the 
catalogues), respectively. If a pair $(i, j)$ has $n_g\geq 10$ and $\Delta_f > 1$, we consider it is linked by a filament.

We identify $441$ Nodes and $449$ Filaments inside the DR16+ flux-limited sample.


\subsubsection{The sample of Loose Groups}
\label{sect:lg_sam}

From the sample of FoF galaxy groups, we select as Loose Groups (hereafter LGs) those that have not been targeted as Nodes nor CGs. 

This sample comprises $13\,812$ LGs having four or more galaxy members. 

\subsection{The sample of Cosmic Voids}
\label{sect:iden_voids}

We identify cosmic Voids following the selection procedure described by \cite{ruiz+15} from a volume-limited sample of galaxies built from DR16+. The identification algorithm for an observational sample consists of the following steps \citep{ruiz+19}:

\begin{description}
    \item[(i)] We perform a Voronoi fragmentation \footnote{The Voronoi diagrams are computed using the public library \textsc{voro++} of \citealt{2009Chaos..19d1111R}.}, which requires defining a region of space around a given galaxy, where any point inside the region is closer to that galaxy than to any other.
    Hence, the density cell is estimated as the inverse of the volume of the Voronoi cell ($\rho_{\rm cell}= 1/V_{\rm cell}$) and the density contrast is defined as $\delta~=~ \rho_{\rm cell}/\bar{\rho} - 1$, where $\bar{\rho}$ is the mean density of tracers.

    \item[(ii)] Candidate regions for Voids are centred at the position of cells whose density contrast satisfies $ \delta < - 0.7$.
    
    \item[(iii)] From the candidate centres, the integrated density contrast $\Delta$ is calculated iteratively  within spheres of increasing radius $R$ as follows:     
    $$\Delta(R)=  \frac{3}{R^3} \int_{0}^{R}\delta(r) r^2dr$$
    When the integrated overdensity satisfies $\Delta(R)< - 0.9$, the iteration ceases and the radius of the current sphere is defined as the radius of the Void candidate ($R=R_{\rm void}$). If the threshold $\Delta(R_{\rm void})$ is never reached, the candidate is discarded.
    
    \item[(iv)] To define the regions found as candidate Voids, it is necessary to select as best as possible the radius and centre of these objects. To achieve this, step (iii) is executed recursively, starting from a randomly shifted centre instead of the previously defined centre of the candidate. This displacement will be random and proportional to the radius of the candidate void. Each displacement is accepted if the new radius obtained is larger than the last value. If the current step is accepted, the centre of the candidate is updated to the new position. This procedure provides a well-defined centre and a maximum radius.

    \item[(v)] Finally, all overlapping spheres are rejected, taking the candidate with the largest radius of $R_{\rm void}$.
    
\end{description}

We use galaxies from the V1 sample as tracers to identify spherical cosmic Voids and select cosmic Voids that have at least $80$ per cent of their volume within the catalogue. The final sample comprises $659$ Voids with $3\,493$ galaxy members. The distribution of void radii spans from $6.06$ to $22.21 \ \Mpc  \, \rm h^{-1}$ with a median of $9.77 \, \Mpc \, \rm h^{-1} $.

Following \cite{2013MNRAS.434.1435C}, it is possible to define two subsamples of Voids according to their dynamics and the surrounding environment, S-type (for the word ``Shell'') and R-type (for the word ``Rising'') Voids. 
The former are Voids immersed in overdense environments with respect to the mean density of the Universe, which in the future will collapse gravitationally due to the overdense wall surrounding them. 
The second type consists of Voids immersed in subdense environments, which can be treated as isolated regions in isotropic expansion. 
The distinction between these two types is made on the basis of the profile of $\Delta (r)$ in a ring of 2 to 3 void radii, $\Delta_{2-3}(r)$. Therefore, we select those Voids with $\Delta_{2-3}(r) > 0$ as S-type Voids and Voids with $\Delta_{2-3}(r) < 0$ as R-type Voids. The subsamples consist of $406$ S-type Voids comprising $1\,802$ galaxies and $253$ R-type Voids with $1\,691$ galaxy members.
The distribution of the radii of the samples defined by their dynamics are within $6.14 < R_{\rm void} < 21.25$ \Mpc $\,\rm h^{-1}$ for S-type and  $6.06 < R_{\rm void} < 22.21$ \Mpc  $\,\rm h^{-1}$ for R-type, with medians of $9.42$ and $10.29$, respectively.



\section{CGs in different environments}
\label{sec:cg_env}

The identification of structures in DR16+ described in the previous section determines only the cosmological structures involved in this work.
The next step is to associate CGs with these structures. 

We will classify CGs that are associated either with Nodes of filaments ($CG_N$), Filaments ($CG_F$), Loose groups ($CG_L$), both types of Voids ($CG_{VS}$ and $CG_{VR}$), or not belonging to any of these structures, CGs Non-Embedded ($CG_{NE}$). 

\subsection{CGs in Nodes}
\label{sec:cg_in_fof}

First of all, we select those CGs that can be considered embedded in Nodes of filaments. To do this, we perform a member-to-member comparison choosing as CGs in Nodes those CGs that share at least one member with the Node. If a CG is associated with more than one Node, we choose the one that shares the most members.


Following the procedure described above, we find that $186$ CGs can be classified as embedded in Nodes of filaments, with $860$ galaxy members.

\subsection{CGs in Filaments}
\label{sec:cg_in_fil}

As we mentioned in Sect. \ref{sect:fil_sam}, the method to identify filamentary structures associates galaxies to the Filaments within a cylinder in redshift space with length given by the separation between Nodes, and a given fixed radius. If a CG centre is within the region delimited by the Filament, then it is considered as a CG embedded in a Filament ($CG_F$). Also, we perform a member-to-member comparison between the members of CGs and Filaments and complete the sample of CGs in Filaments by adding those CGs that share at least one member with the Filament.

Only CGs not embedded in Nodes (previous section) are cross-matched with the Filaments. The sample of CGs in Filaments comprises $61$ CGs with $275$ galaxy members. 

\subsection{CGs in Voids}
\label{sect:cgs_voids}

To study the location of each CG with respect to the cosmic Voids, the spatial distances between the centres have been calculated. 
In this case, the centres of the CGs correspond to the centres of the smallest circle circumscribing the centres of CG member galaxies, and the centres of the Voids are the geometric centres of the sphere that defines the void with $\Delta(R_{\rm void}) <  - 0.9.$
The position vectors ($x_{i}$, $y_{i}$, $z_{i}$) of both systems centres are determined based on the equatorial coordinates ($\alpha$, $\delta$) and the comoving distances $d_{\rm com}$ computed from the redshifts of the objects.
For each CG not embedded in Nodes or Filaments, we calculate the 3D distances to all cosmic Voids. Then, we normalise the distances to the radius of each Void, and determine the minimum normalised distance, $D^N_{\rm min}$, i.e., for each CG we find its nearest cosmic Void.
We adopt a threshold for the minimum normalised distance to define when a CG inhabits a Void: if $D^N_{\rm min} < 1.1$ the CG is embedded in a cosmic void. Also, we extend the sample of CGs in Voids adding those CGs that share members with the galaxies members (including faint galaxies) of Voids, although the centre of CGs might be out of the boundaries of the Voids. Therefore, we consider CGs as embedded in Voids if they have at least one common member or if the centre of the CG fulfils that  $D^N_{\rm min} < 1.1$. We find that $70$ CGs can be considered associated with Voids: $48$ CGs with S-type Voids ($CG_{VS}$) comprising  $167$ galaxy members, and $22$ in R-type Voids ($CG_{VR}$) with $75$ galaxies.



\subsection{CGs in Loose Groups}
\label{sec:cg_in_LG}

For the remaining CGs, i.e., those that belong neither to Nodes, Filaments nor Voids, 
we analyse if they can be associated with LGs.
The association was also determined according to a member-to-member comparison. We consider a CG embedded in an LG if they have at least two common members.
According to this criteria we find that $436$ CGs can be considered embedded in LGs, which contain $1\,706$ galaxy members.


\subsection{Non-Embedded CGs}
\label{sec:cg_NE}

The sample of Non-Embedded CGs comprises all those CGs that are not associated with any of the previously detailed cosmological structures (Nodes, Filaments, cosmic Voids, LGs). This sample may include CGs that are relatively isolated, as well as CGs that could be formed by pairs of galaxies, or even CGs embedded in loose groups with lower overdensity contrast than the sample used in this work or in Filaments with less than 10 members. 

This sample comprises $615$ CGs with $2\,468$ galaxy members. 
Therefore, we find that $45 \pm 3$ per cent of the total sample of CGs in the volume-limited sample will be considered Non-Embedded systems.


\section{Results}
\label{sec:results}

We adopt two different approaches to study the impact of the environment on CGs. 
Firstly, we study the properties of CGs as a function of the environment in which they reside. Secondly, we study the properties of galaxies in CGs depending on the different environments that surround them.
In Table \ref{tab:N_cg} we show the number and percentages of CGs associated with each environment.

\begin{table}
\centering
\caption{Number of CGs in different environments. The errors were calculated using the 95 per cent binomial confidence interval for each percentage computed as $\pm 1.96 \sqrt{f(1-f)/N}$, where $f$ is the fraction of CGs per environment, and $N$ is the total number of CGs.}
\begin{tabular}{lll|rr}
\hline
Samples & Environments&&    $\#$ CGs &  Percentage\\
\hline 
&All & &1368&\\
\hline           
$CG_N$& Nodes of filaments   &    & $186$ & $14 \pm 2$ \%\\
$CG_F$& Filaments           &    & $61$ & $4  \pm 1$  \%\\
$CG_{VS}$&   Voids S-Type && $48$  & $3 \pm 1$  \%\\
$CG_{VR}$&   Voids R-Type && $22$   & $2 \pm 1$  \%\\
$CG_{LG}$& Loose groups &        & $436$ & $32  \pm 2$ \%\\ 
$CG_{NE}$& CGs Non-Embedded          &    & $615$  & $45 \pm 3$ \%\\
\hline
\end{tabular}
\label{tab:N_cg}
\end{table}
\begin{figure}
    \centering
    \includegraphics[width=0.48\textwidth]{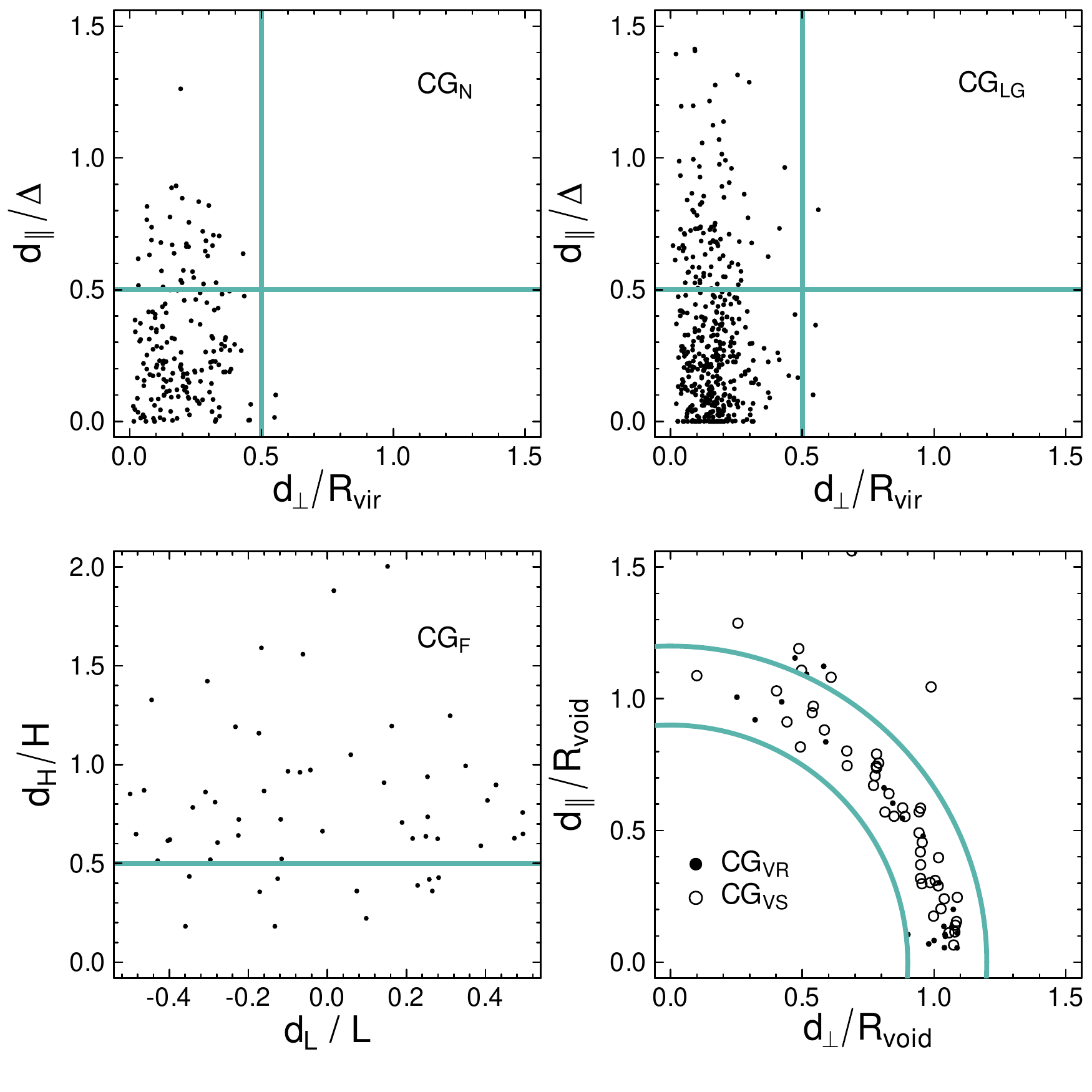}
    \caption{Projections of the 3D comoving distance between the CG centre and its host system normalised to the characteristic sizes of the hosts. For CGs in Nodes (top left), LGs (top right) and Voids (bottom right), the projections are along the line of sight and on the sky plane. For CGs in Filaments (bottom left), the projections are on the filament axis and in the direction perpendicular to the axis.
    }
    \label{fig:pos_cg}
\end{figure}

As we described in Sect.~\ref{sect:fil_sam}, Nodes of filaments are massive FoF groups. We find that $14 \pm 2$ per cent of CGs inhabit Nodes while $32 \pm 2$ per cent are in LGs (see Table~\ref{tab:N_cg}). Therefore, $46 \pm 4$ per cent of CGs are embedded in FoF Groups.
In contrast, \cite{diaz15} found that $27\pm 5$ per cent of their sample was embedded in FoF groups, a lower percentage than ours.
They identified groups in the K-band with four or more members and used the {\it classic} algorithm to identify CGs. \cite{Taverna+16} demonstrated that identifying CGs in different bands produces differences in the resulting CG samples. In addition, in this work we are using the {\it modified} algorithm to identify CGs, which identifies twice more CGs than the classic algorithm \citep{DiazGimenez+18}. This last point and the addition of triplets to our sample may be the main causes of the increase in the percentage of CGs embedded in FoF groups in our sample. 
On the other hand, \cite{mendel+11} found $\sim 50$ per cent of CGs are isolated and the other half are CGs embedded in rich clusters. 
\cite{Zheng+21} found that $\sim 27$ per cent are isolated CGs and $\sim 26$ per cent are CGs embedded in systems but they dominate the luminosity of
the halos (we adopted the latest category as Non-Embedded systems in this work), and $\sim 23$ per cent are CGs embedded in large clusters where the CG does not dominate the luminosity of their parent group, while \cite{Sohn+16} found that only $\sim 23$ per cent of CGs are in dense environments. 

Despite the different selection criteria among CG samples that make it difficult to perform a fair comparison of percentages of embedded systems, it seems that the percentage of CGs that can be considered roughly isolated (relatively free from external conditioning) is considerably high. 
%
We find that $\sim 50$ per cent of CGs are inhabiting high-density regions in the Universe (Nodes, Filaments and LGs), while the remaining $\sim 50$ per cent are preferentially located in low-density regions. Nevertheless, only a small percentage of those CGs are located in the lowest density regions of the Universe, i.e, only $5$ per cent CGs are associated with Voids, while the remaining large fraction of CGs ($45$ per cent) are found roaming in regions as dense as the mean density of the Universe. 

\subsection{Location of CGs within larger structures}
To analyse the position of CGs embedded in different structures, we compute the 3D comoving distance between the CG centre and the centre of its host system.

The definition of CGs in redshift space with cylinders of size $\pm 1\,000 \, \rm km/s$ might allow the CGs to be quite scattered along the line-of-sight within the systems they inhabit. In addition, given the different geometries and different methods for identifying structures, to show the results we use different projections of the 3D comoving distance between the centres depending on the environment in which the CGs are inhabiting. 
For CGs embedded in Nodes, Loose Groups and Voids, we use the projections along the line-of-sight ($d_\parallel$), and on the plane of the sky ($d_\perp$). For CGs embedded in Filaments, we use the projection along the axis that joins the Nodes centre, $d_L$, and the projection perpendicular to the filament axis, $d_H$. 

The scatter plots of the distances in the two directions are shown in Fig.~\ref{fig:pos_cg}. We normalise the projections to the characteristic sizes of the host systems. 
For CGs in Nodes and Loose groups: the projection along the line-of-sight is normalised by half the maximum separation among galaxies in the host system ($\Delta$), while the projection on the plane of sky is normalised by the virial radius of the host ($R_{\rm vir}$). 
For CGs in Voids, the projections are normalised to the radius of the host Voids ($R_{\rm void}$). And for CGs in Filaments, the projection along the axis of the filament is normalised by the length of the filament ($L$, distance between Nodes), while the other direction is normalised by the fixed radius of the filament ($H=1.5 \, \rm Mpc \, h^{-1}$).

We split the sample of CGs into two according to the location of CGs in their hosts: those inhabiting the inner parts of the host, and those in the outer regions. For Nodes, LGs and Voids, we consider the inner parts the region within half the virial radius and half the size of the host along the line of sight. For Filaments, the inner region is defined as closer to the axis of the Filament, i.e, where the perpendicular distance to the axis is half the radius of the filament. In Fig.~\ref{fig:pos_cg} the inner regions of the systems lie below the cyan solid lines. 
We find that $78$ per cent of CGs in Nodes, $76$ per cent of CGs in LGs, and $18$ per cent of CGs in Filaments lie in the inner regions of their hosts. None of the CGs lie in the inner regions of Voids; instead they lie in the spherical shell surrounding Voids, most of them 
are located between $\sim 0.9$ and $1.2$ times the Void radius (shown as circles in the figure Fig.~\ref{fig:pos_cg}).



\begin{figure}
    \centering
    \includegraphics[width=\columnwidth]{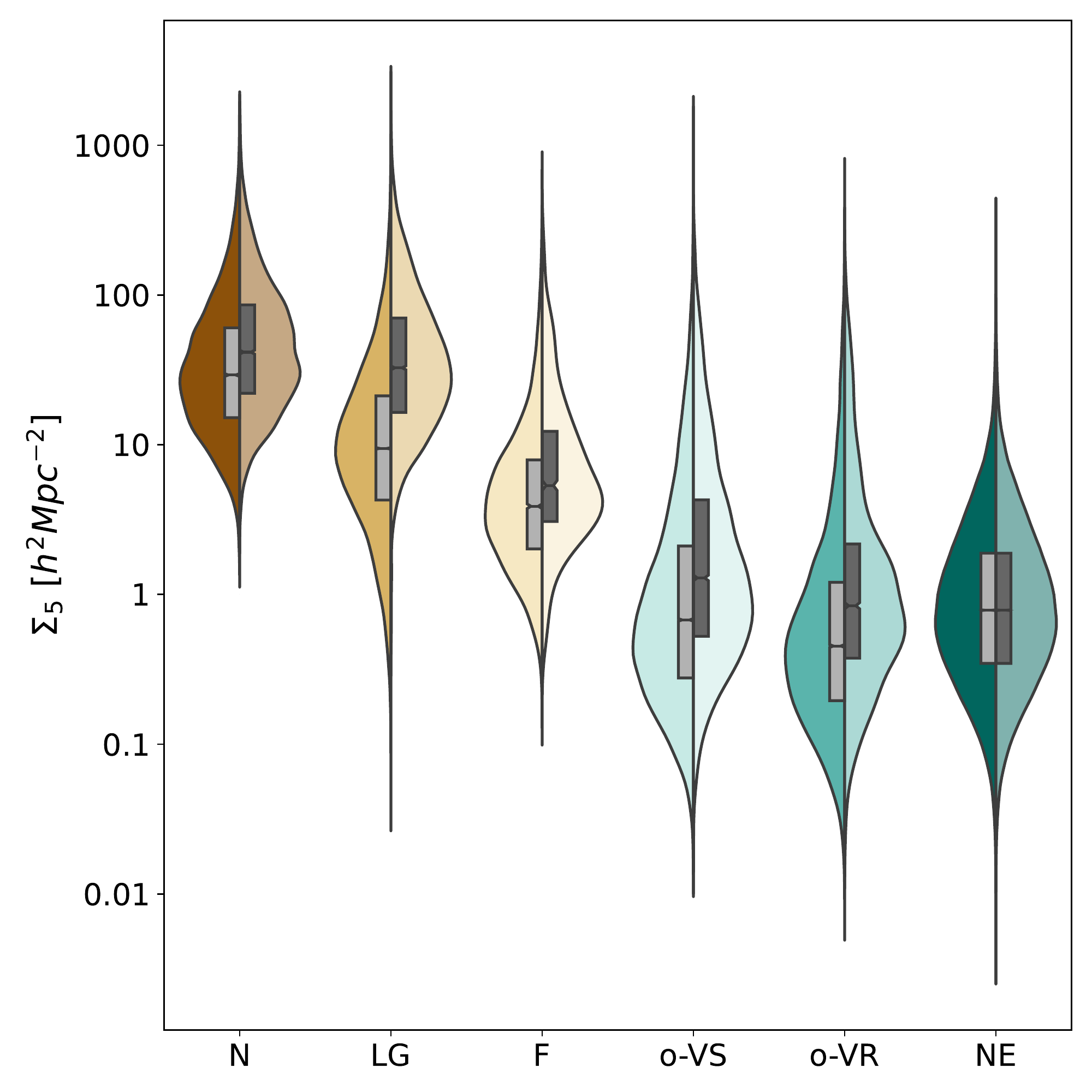}
    \caption{Distributions and boxplots of the projected local density of galaxies to their fifth nearest neighbour, $\Sigma_5$, for galaxies inhabiting the different structures in V1 used in this work 
    (N: Nodes, LG: Loose Groups, F: Filaments, and o-V: the outer nearby region around Voids), as well as for galaxies that are not embedded (NE) in any of those structures. Left distributions correspond to galaxies in all the corresponding structures, while right distributions correspond to galaxies only in those structures that host CGs (except for galaxies in Non-Embedded environments in which both distributions are the same).  
    }
    \label{fig:den5}
\end{figure}

\subsubsection{Local density}
\label{sec:local_den}
We examine the local density of galaxies in each environment as a proxy for characterising each environment.

First of all, we restrict the samples of Nodes, Filaments and LGs to those within the volume-limited sample, V1.  As done with CGs, we keep those systems whose brightest galaxy satisfies  $M_{\rm bri} \le -19.769$ and $z_{\rm bri} \le 0.1$. This restriction allows systems to have other fainter galaxies. 

The sample of Nodes in V1 comprises $431$ systems with $11\,002$ galaxies, while the sample of Filaments has $313$ Filaments with $3\,781$ galaxies. In the case of LGs, in this section, we have not included those LGs that, in turn, are inhabiting Filaments or Voids. Therefore, we end up with $13\,201$ LGs with $50\,876$ galaxies. 
Voids have already been identified in V1. As a result of the previous section, we observed that most CGs in Voids lie in the outer neighbourhood of Voids. Therefore, in this section, we select galaxies in  DR16+ flux-limited sample that are in the outer regions of Voids (with 3D comoving distances between $0.9$ and $1.2 R_{\rm void}$). There are $28\,351$ and $18\,225$ galaxies outside Voids S and R, respectively.
We have also included galaxies considered not embedded in any of the structures defined in this work. In this case, we only imposed the redshift restriction to the galaxies to select $215\,848$ objects up to $z~\leq~0.1$. As mentioned before, this sample might include pairs of galaxies, or even galaxies in loose groups with lower overdensity contrast than the sample used in this work, galaxies in Filaments with less than 10 members, as well as galaxies inhabiting relatively dense regions such as isotropic infall regions.

For each galaxy in the structures defined above, we compute the local density by using the projected distance to its 5th nearest neighbour in the DR16+ flux-limited sample that lies within $1000 \ \rm km/s$ from the galaxy, $\displaystyle \Sigma_5 = 5/(\pi \, r_5^2)$. 
In addition, we select those Nodes, Filaments, Voids and LGs that are hosting CGs. Therefore, we compare the local density of galaxies in the different structures with the local density of galaxies inhabiting those structures that host CGs.  

Figure~\ref{fig:den5} shows the distributions of local densities for galaxies within each structure.  The distributions of local densities of galaxies in all the identified systems are shown in the left half of the plots, while the distributions for galaxies in those systems hosting CGs are shown in the right halves. For Non-Embedded galaxies, we have just repeated the distributions in both halves. 

Considering the distributions on the left (all systems), although the distributions overlap, we observe that the medians of the distributions are statistically different between the different samples. 

Something interesting happens when analysing only those systems hosting CGs (right halves). The local densities of galaxies in these subsamples tend to be higher than those in the complete samples. The most noticeable differences appear for galaxies in LGs and in the outer regions of Voids. The local densities of galaxies in LGs that host CGs are mostly at the tail of the distribution of densities of the whole sample of LGs, being much similar to the local density of galaxies in Nodes.  On the other hand, the local density of galaxies in the outer regions of Voids that host CGs are shifted towards larger values, being slightly higher than the local density of Non-Embedded galaxies. 

Hereafter, we decided to order the different structures according to the local density of their galaxies when they are hosting CGs, i.e., in descending order:  Nodes, LGs, Filaments, outer regions of Voids and the Field. 

\subsection{Influence of environments on CG properties}
\begin{figure*}
\begin{center}
\includegraphics[width=0.99\textwidth]{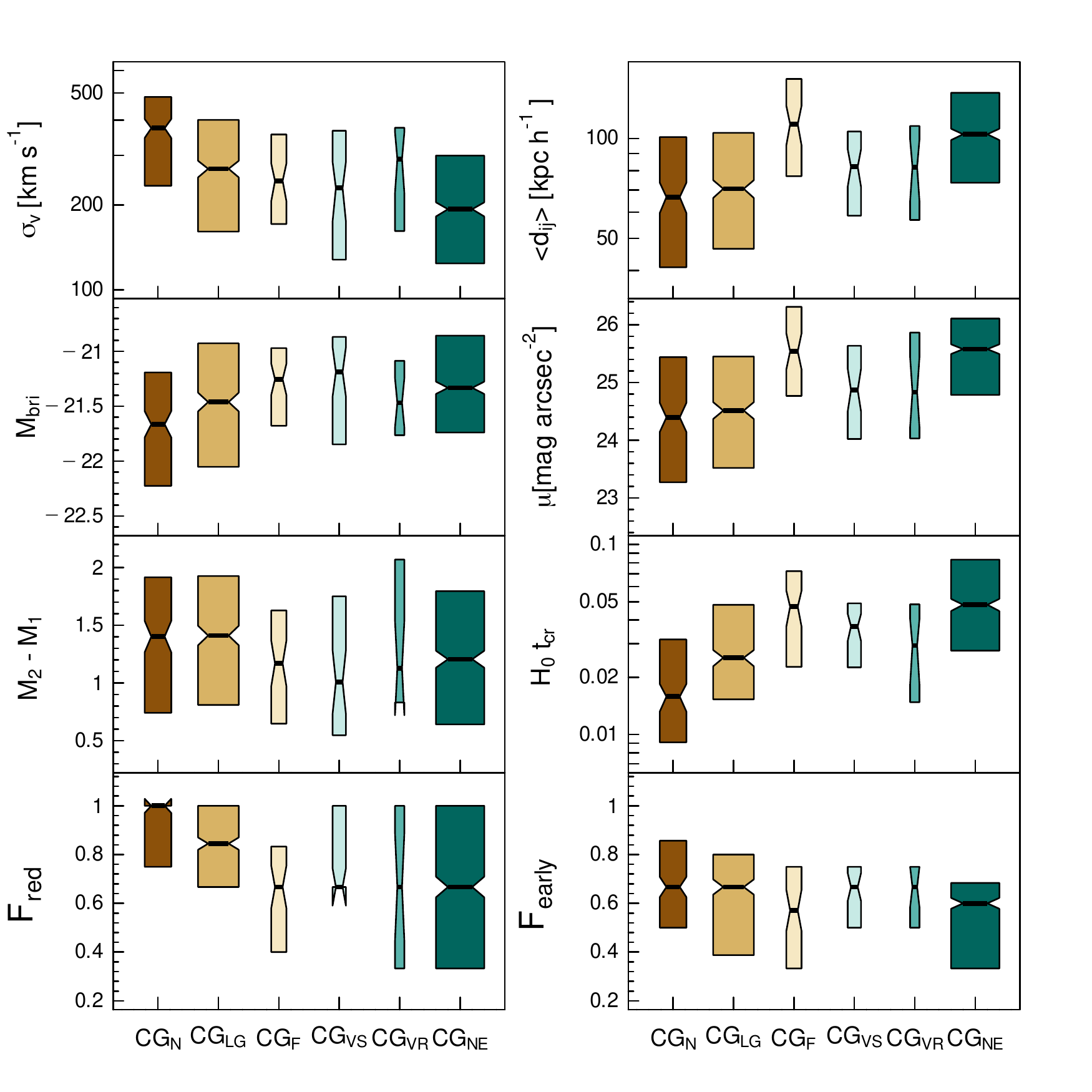}
\caption{Boxplot diagrams of properties of compact groups inside different environments: Nodes ($CG_N$), LGs ($CG_L$), Filaments ($CG_{F}$), S-type Voids ($CG_{VS}$) and in R-type Voids ($CG_{VR}$) and those that can be considered Non-Embedded $(CG_{NE})$. The box extends from the lower quartile (25th percentile) to the upper quartile (75th percentile). The notches indicate the approximate 95 per cent confidence interval for the medians.} 
\label{fig:boxplot}
\end{center}
\end{figure*}


\begin{figure*}
\begin{center}
\centering
\includegraphics[width=1.9\columnwidth]{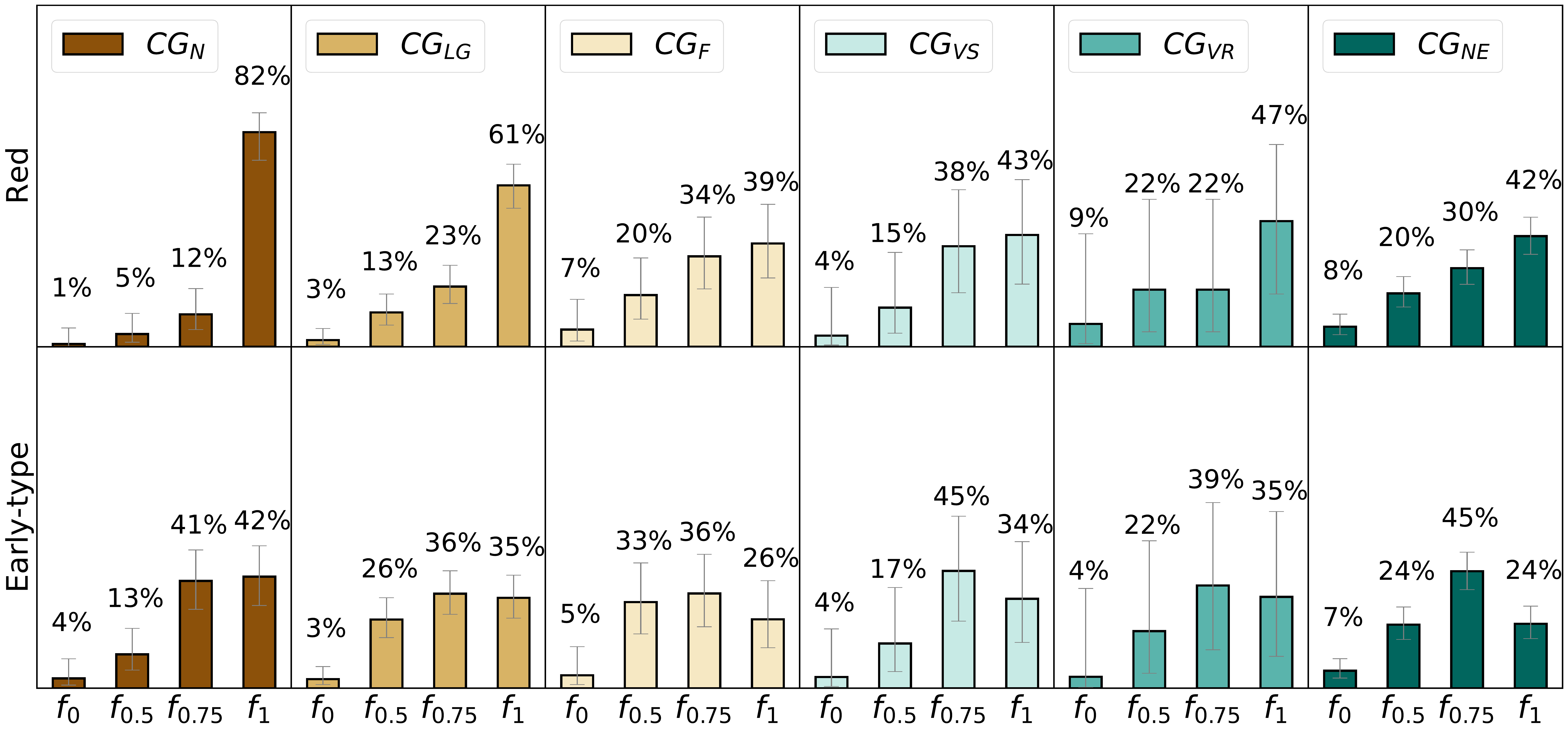} 
\caption{
Percentages of CGs in subsamples split according to their fraction of red ({\it upper panels}) or early-type ({\it lower panels}) galaxies.
The subsamples comprise CGs whose red/early-type fractions are within the following ranges: $f_0=[0,0.25)$, $f_{0.5}=[0.25,0.5)$, $f_{0.75}=[0.5,0.75)$, and $f_{1}=[0.75,1]$. The error bars are the binomial 95 per cent  confidence interval estimated using the Wilson score interval \citep{wilson27}.
}
\label{fig:barplots}
\end{center}
\end{figure*}

In this section, we study the effects of the environment on the main properties of CGs. All properties of CGs were calculated following the same procedure described in \cite{zandivarez+22}, but we use as CGs members both the bright members and the faint galaxies in the region. 
We focus our study on the following CG properties: 
\begin{itemize}
    \item $\sigma_v$: Radial velocity dispersion, calculated using the gapper estimator described by \citet{Beers90}.
    \item $M_{\textup{bri}}$: r-band rest-frame absolute magnitude of the brightest galaxy member.
    \item $\Delta M_{12}$: Absolute magnitude difference between the two brightest galaxies.
    \item $\langle d_{ij} \rangle$: Median of projected separations among galaxy members.
    \item $\mu$: r-band mean group surface brightness.
    \item $H_0 \, t_{cr}$: Dimensionless crossing time.    
\end{itemize}

In Fig.~\ref{fig:boxplot} we show the boxplot diagrams of CG properties for CGs split according to the environment they inhabit. 
Two samples are statistically different if the notches of the boxes (confidence intervals of the medians) do not overlap. We find significant differences among environments in some of the properties shown.

The radial velocity dispersion shows interesting behaviour. 
CGs in Nodes have the highest median values of $\sigma_v$ and they are followed in descending order by CGs in LGs, Filaments and in the surroundings of  Voids reaching the smallest median for Non-Embedded CGs.
The median velocity dispersion observed for CGs in Nodes almost doubles that observed for Non-Embedded CGs in the lowest-density environments.
In other words, CG velocity dispersion increases with the density of the environment they inhabit.

In terms of the luminosity of their brightest galaxy, we note that the CGs in Nodes and LGs have the brightest first-ranked galaxies while there is no clear difference among the remaining environments. 
When analysing the dominance of the brightest galaxy ($\Delta M_{12}$), we observe a similar behaviour as previously described for the first-ranked galaxies. CGs in Nodes and LGs show the largest magnitude gap ($\sim 1.4$), while the remaining environments display smaller gaps ($\sim 1$). 

When analysing the properties related to CG sizes, the median of the inter-galaxy projected separation $\langle d_{ij} \rangle$ of CGs in Nodes and in LGs are the smallest, followed by larger values for CGs in the surrounding Voids. CGs in Filaments and Non-Embedded CGs display the largest values of $\langle d_{ij} \rangle$. 
The distributions of group surface brightness (which is related to the compactness of the systems) closely resemble the size distributions. 
The adimensional crossing time distributions are, by construction, a mixture of the results obtained for the radial velocity dispersion and group sizes. Galaxies in Non-Embedded CGs or CG in Filaments need approximately $\sim 0.7$ Gyrs (in median) to get across the system, while their counterparts inhabiting CGs in Nodes need roughly a third of that time.

These results to a certain extent reinforce the idea that the global environment around CGs shapes their internal evolution since the denser the environment CGs inhabit, the larger the group velocity dispersion, the brighter the first-ranked galaxy, the smaller the group size, the higher the compactness and the smaller the crossing time, 
making them more prone to experience galaxy interactions.

\subsection{Influence of environments on galaxies in CGs}

We also study the fraction of red and early-type galaxies that inhabit each CG. We followed the methodology used by \cite{zandivarez11} to classify galaxies as red/blue accordingly to whether their $u-r$ colour is larger/smaller than the luminosity-dependent relation for the galaxy colour\footnote{The empirical relation is $P(x)= -0.03077 x^2 - 1.4074 x - 13.64045$, where $x =M_r - 5\, \log{h}$} (see Appendix~\ref{ap:pasivas} for a detailed description). 
Additionally, we estimate the fraction of early-type using the concentration index parameter $C=r_{90}/r_{50}>2.6$ following \cite{Strateva01} (see Appendix~\ref{ap:early-type} for details). 

In the bottom panels of Fig.~\ref{fig:boxplot}, we show the boxplot diagrams of the distributions of the fraction of red galaxies (bottom left panel) and the fraction of early-type galaxies (bottom right panel) for CGs in each environment.

Most of the CGs in Nodes are clearly dominated by a red galaxy population, while CGs in LGs also show a very large fraction of red galaxies ($\sim 0.85$ in median). CGs in Filaments, Voids and those considered Non-Embedded have a relatively high median of the fraction of red galaxies ($\sim 0.67$). Particularly, the fraction of red galaxies in CGs in S-Voids shows positive skew (i.e., the tail of the distribution tends to higher fractions compared with the median).

When observing the fraction of early-type galaxies in CGs (bottom right panel of Fig.~\ref{fig:boxplot}), there are no very noticeable differences in the medians as a function of environments.
All samples of CGs have a similar median of the fraction of early-type galaxies, which ranges between $0.6$ and $0.7$. 
The only noticeable feature in the distributions of the fraction of early-type galaxies is that 
all but the fraction of CGs in Nodes display a distribution with negative skew. 

To deepen our analysis of the type of galaxies that populate CGs, 
we split the samples of CGs inhabiting each environment into four subsamples according to their fraction of red(early-type) galaxies: the first subsample comprises CGs that have less than $25$ per cent of their member galaxies being red (early)-type ($0 \le \rm Fraction < 0.25$); the second subsample comprises CGs whose fraction of red (early-type) galaxies is within the range $[0.25,0.5)$; the third, with fractions within $[0.5,0.75)$; while the fourth subsample comprises CGs with most of their members being red (early-type) galaxies (i,e: fraction within  $[0.75,\ 1]$). We refer to these sub-samples as  $f_0$, $f_{0.5}$, $f_{0.75}$, and $f_1$, respectively.
  
Top panels of Fig.~\ref{fig:barplots} show 
the percentages of CGs within each subsample that are split according to their fraction of red galaxies, while the bottom panels show the percentages when splitting the CGs according to their fraction of early-type galaxies. From left to right, we show the samples of CGs inhabiting the different environments.

When analysing the subsamples split by their fraction of red galaxies (upper panels), there are very few CGs in the first two intervals, i.e: most CGs have at least half of their members being red galaxies, regardless of the environment they inhabit. In addition, the vast majority ($82$ per cent) of the CGs in Nodes show colour concordance of red galaxies (i.e., they are formed with at least three-quarters of their members being red galaxies), while this percentage diminishes towards less dense environments. 
Roughly $61$ per cent of CGs in LGs have at least $75$ per cent  of red galaxies, while the other less dense environments show less than $50$ per cent of CGs with this high proportion of the red population. The fraction of CGs with a colour concordance of blue galaxies, $f_0$, in all the environments is very low. 

When splitting the CGs according to their fraction of early-type galaxies (bottom panels), we observe that $83$ per cent of CGs in Nodes have more than $50$ per cent of early-type galaxies. For the other environments, this percentage is between $70-79$ per cent, except for CGs in Filaments where only $62$ per cent of CGs have more than half of their members being early-type. 
Approximately $42$ per cent of CGs in Nodes show morphological concordance (fraction of early-type galaxies greater than 0.75), while for the rest of the environments the percentage of CGs is lower (the lowest values are found in Filaments and Non-Embedded CGs).  

These results are in agreement with the expected behaviour based on our previous results. CGs inhabiting the densest regions in the Universe (such as Nodes) are small systems with a considerably agitated dynamic, a very bright central galaxy and its membership is dominated by red galaxies where several of them can be considered as early-type. At the other end of the density spectrum, CGs in the most isolated regions in the Universe are relatively larger systems formed by slow-moving galaxies, with red galaxies that barely dominate over the blue population. Therefore, although CGs were all selected in the same way, their nature and evolution could be conditioned by the environment they inhabit.

From the work of \cite{mendel+11} it can be inferred that the $\sim 60$ per cent of isolated CGs show a colour concordance of red galaxies (i.e., more than 75 per cent of their members have red colours), and $\sim 90$ per cent of them have more than 75 per cent of early-type members. These percentages are larger than those we observed for Non-Embedded CGs (42 per cent and 24 per cent for red and early-type galaxies, respectively). 
For embedded CGs in rich structures, they found that $\sim 90$ per cent of them display a red colour concordance and the same percentage of CGs with the highest rate of early-type galaxies. We only observed this large percentage of red galaxies in CGs in Nodes, while none of the structures used in this work display such a high fraction of CGs with a high fraction of early-type galaxies. Nevertheless, a comparison between these two works is rather difficult due to the different criteria to identify CGs and select embedded and isolated CGs.

\begin{figure*}
\begin{minipage}[b]{0.8\linewidth}
\centering
\includegraphics[width=\linewidth]{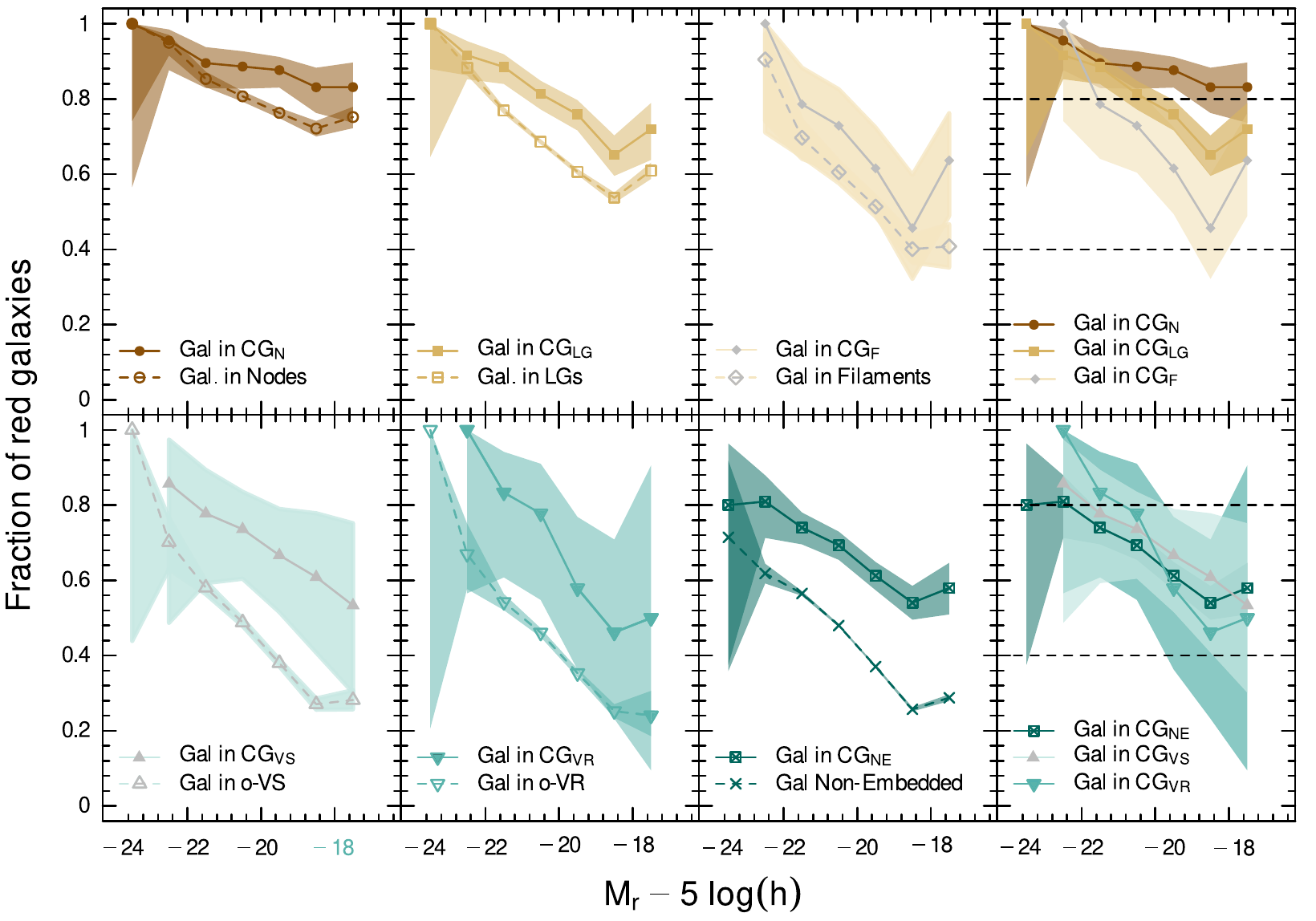}
\includegraphics[width=\linewidth]{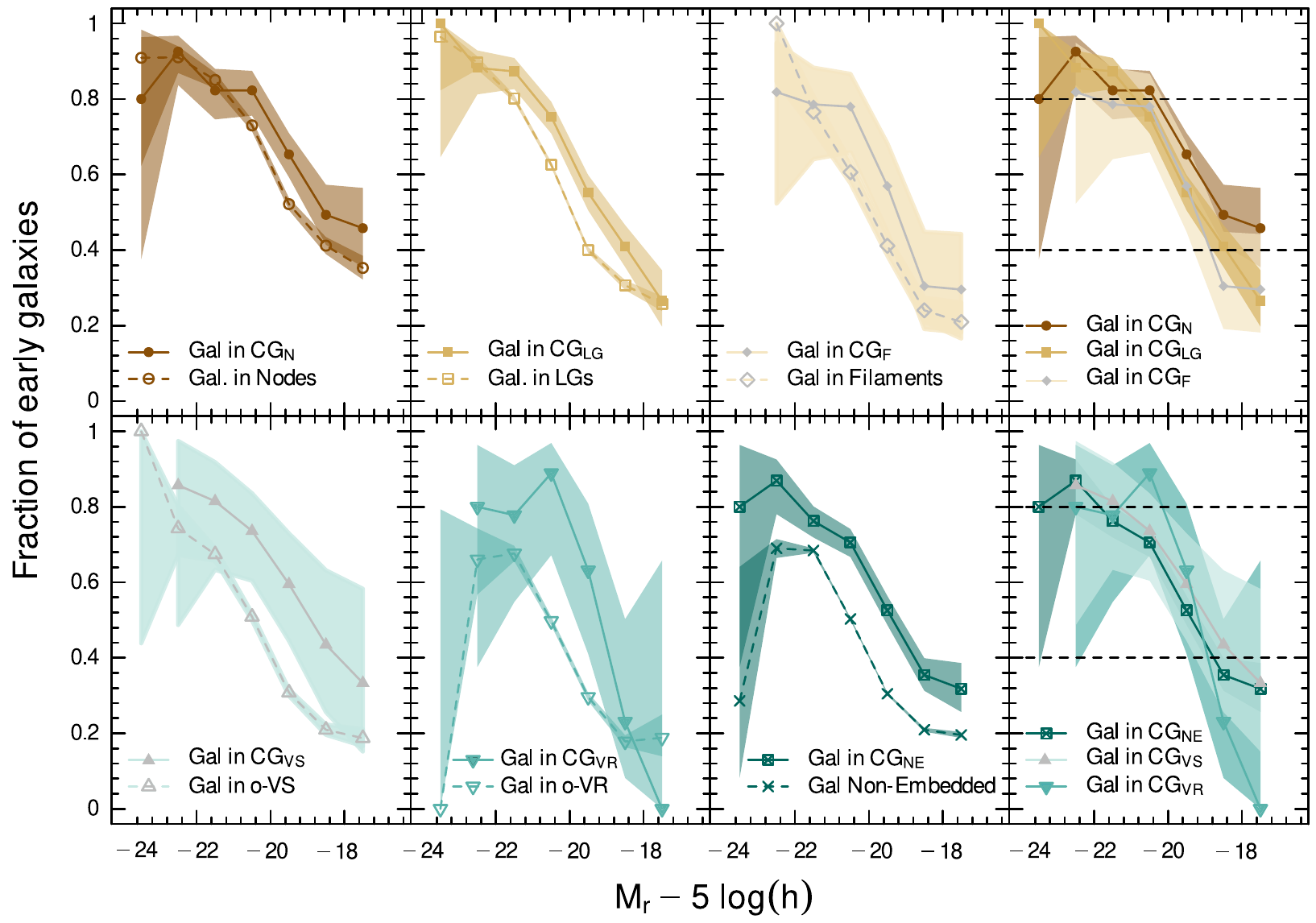}
\caption{Fractions of red ({\it upper panels}) and early-type ({\it bottom panels}) galaxies per bin of absolute magnitude. Solid lines correspond to galaxies in CGs embedded in different environments (different boxes), while dashed lines correspond to galaxies within the corresponding environment, regardless of being in CGs. Shaded areas correspond to the binomial errors computed for the fractions.
In the {\it rightmost panels} we only show the fractions of red/early-type galaxies in CGs inhabiting the different environments. Horizontal dashed lines are for comparison purposes only.}
\label{fig:red_frac_pass}
\end{minipage}
\end{figure*}


\subsubsection{Environmental impact on galaxies: Local vs Global}

We study the fraction of red and early-type galaxies in CGs embedded in different environments, this time as a function of their absolute magnitudes. 
For comparison, we also analyse these fractions for galaxies that inhabit the same type of host environment. 
In order to make a fair comparison, galaxies in the different environments have been selected in such a way that the brightest galaxy of the structure is within the volume-limited sample V1, except for galaxies that do not belong to any structure where only the redshift restriction was applied to all galaxies. These galaxy control samples have been described in Sect.~\ref{sec:local_den} and used to compute the local densities. 

We split the samples of galaxies into seven equal bins of absolute magnitudes (1 mag size bin) and measure the fractions of red/early-type galaxies per bin.  
These trends are shown in Fig.~\ref{fig:red_frac_pass}. The upper plot corresponds to the fractions of red galaxies, while the bottom plot shows the fractions of early-type galaxies. The fractions of galaxies in CGs are shown as solid lines, while the fractions of galaxies in the corresponding  environments are shown in dashed lines. 
Each panel corresponds to a different host environment (see inset legends). For a better comparison between galaxies in CGs embedded in different environments, the trends (solid lines) are re-arranged in the rightmost panels. The horizontal dashed lines are for comparison purposes only. The fraction errors are the binomial 95 per cent  confidence interval estimated using the Wilson score interval \citep{wilson27} and they are shown as shaded areas.

Regarding the fraction of red galaxies, the comparison shows that CGs inhabiting a given environment have a higher fraction of red galaxies than expected in such environments. And this stands for most environments (except Filaments) and most bins of absolute magnitudes. Only at the brightest magnitude bins, the fractions of red galaxies are the same for both samples. 
The difference between the fractions of red galaxies in CGs and the fractions of red galaxies in the host environments increases towards the lower-density environments.  

In addition, we find a weak dependence on the absolute magnitudes of the fraction of red galaxies in $CG_{N}$. We observe a variation of only $0.18$ over the entire range of absolute magnitudes. 
However, the fractions of red galaxies in the remaining samples show a strong variation as a function of the absolute magnitude. The fainter the magnitude bin, the lower the fraction of red galaxies, although the fraction of red galaxies in CGs does not fall below $0.5$ in any environment, which does not happen for galaxies in the host environments.  
From the comparison in the rightmost column, the findings of the previous sections that CGs embedded in densest environments (N and LGs) show a larger fraction and higher colour concordance than those embedded in lower density environments can be explained as a consequence of a higher fraction of red galaxies in the fainter magnitude bins, while there is no difference in the fraction of red galaxies in the whole range of magnitudes in CGs embedded in the outer regions of Voids and in the Field, where red galaxies barely dominate among the faintest galaxies.

The bottom panels of Fig. \ref{fig:red_frac_pass} are analogous to the upper panels but for the early-type fraction. In general, we observe a tendency for a higher fraction of early-type galaxies in CGs compared to galaxies in their corresponding environments. Only in the brightest bins, the fractions are indistinguishable. 
In all the environments, we observe a strong dependence of the early-type fractions on the absolute magnitudes, causing the faintest bins to be dominated by late-type galaxies (fractions of early-type $\leq 0.3$). 

In contrast with the red fraction of galaxies in CGs in the rightmost panel, we found no differences among the fraction of early-type galaxies in CGs inhabiting high or low-density environments.

The general trends are in agreement with previous findings by  \cite*{Deng08} and \cite{coenda12} when comparing the fraction of red and early-type galaxies in CGs, LGs and Field. They found that CGs have a larger fraction of  red and early-type galaxies when compared to loose groups and Field galaxies.


\section{Summary and Conclusions}
\label{sec:conclu}
In this work, we performed a detailed analysis about the location of Hickson-like compact groups (CGs) of galaxies in the universe. 
To achieve this goal, we used a sample of CGs recently identified in an extended version of the SDSS DR16 \citep{DR16_sdss} as well as samples of different cosmological structures in the same galaxy parent catalogue: galaxy groups, filaments and nodes, and voids. 

The samples of CGs and galaxy groups were previously identified by \cite{zandivarez+22}, while the samples of filaments and voids have been identified in this work following similar procedures as those described in \cite{Martinez16} and \cite{ruiz+19}, respectively. 
It is worth noticing that the results presented in this work could depend on the methods used to identify galactic structures: CGs, loose groups, nodes, filaments, and cosmological voids. 
Although it is our intention to focus exclusively on the CGs identified with the Hickson criteria, the other structures could be identified in multiple ways\footnote{For comparisons among different finding algorithms see \cite{cautun+18} for voids, and \cite{libeskind+18} for filaments.}. 
A broader comparison involving several methods to define galactic structures could provide insight into the dependence of the results on the particular algorithms used to define environments.
However, the methods adopted here not only are well known in the literature but also our experience at manipulating them allowed us to tailor the resulting samples according to the aims and constraints of the present work.

We adopted different criteria to associate CGs with the cosmological structures: Nodes of filaments, Loose Groups (LGs), Filaments and Voids. Those CGs that cannot be associated with any of these structures have been considered Non-Embedded systems. We restricted the samples of structures to those where the brightest galaxy is within a volume-limited sample given by $M \leq -19.769$ and $z\le0.1$.

To begin with, we found a fifty-fifty chance of finding CGs either embedded in these cosmological structures or not. 
45 per cent  of CGs are not associated with any structure, while a similar percentage of CGs are hosted within Nodes and LGs (46 per cent). 

By analysing the location of CGs within the structures, we observed that CGs are concentrated in the inner parts of Nodes and LGs, mainly in the outskirts of Filaments, and in the outer nearby shell surrounding Voids. To have a better understanding of the structures involved in this study, we study the local density of galaxies within these structures by computing the projected distance to the fifth nearest neighbour. Particularly, we used galaxies in the outer spherical shell around Voids, where CGs are likely to be located. 
As a result, it is possible to visualise the structures in decreasing order according to the local density of the galaxies inhabiting them, and specially in those structures hosting CGs: Nodes, LGs, Filaments, outer regions of Voids and Non-Embedded galaxies.
 
We then focus our attention on the properties of CGs as a function of the structures they inhabit. We observed that in the densest environments such as the Nodes, CGs have the largest velocity dispersions, brightest first-ranked galaxies as well as the smallest sizes and crossing times. 
The opposite behaviour is observed for CGs that can be considered Non-Embedded in any structure used in this work. 
The properties of CGs in LGs are more similar to those of CGs in Nodes, while CGs in Filaments or in the surroundings of Voids display properties mostly in between the two extremes. These results seem to indicate that the density of the environment in which the CGs are immersed has a role to play in their resulting physical properties.

Small crossing times of CGs in Nodes indicate that galaxies need less time to orbit within the system, which would benefit from more interaction between the members. Such interactions could result in mergers that would lead to an increase in the luminosity of the first ranked galaxy, and one would expect that the magnitude difference between the first and the second-ranked galaxy in groups was inversely correlated to the crossing time. We observed a very small (but significant) difference between the median magnitude gaps of CGs in Nodes and LGs ($\sim 1.4$) with those observed in CGs Non-Embedded ($\sim 1.2$). These results are in agreement with those of \cite{Sohn+15} where they have found that the median crossing time of CGs embedded in dense environments is shorter than that of those considered isolated.

Regarding the CG galaxy members, we split galaxies according to properties related to their current star formation and their morphology/shape: galaxy colour and concentration index. We computed the fraction of red and early-type galaxies and analysed the influence of the environment they inhabit. Our results indicate that CGs in Nodes and LGs are dominated by red galaxies, while the other environments show a median of $\sim 60$ per cent of red members. On the other hand, the early-type fractions are rather similar among the different types of environments (60-67 per cent), with some small tendency to be higher for the densest structures. 
\cite{Sohn+16} obtained for a sample of CGs identified with a FoF algorithm in the SDSS DR12 that the fraction of early-type galaxies was 79 per cent in high-density regions and 60 per cent in low-density regions (where density was inferred using the number of nearest neighbours as greater or lesser than 7 neighbours within a given area).
We also observed that 82 per cent of CGs in Nodes exhibit red colour concordance, while 61 per cent of CGs in LGs do. The remaining environments have less than 50 per cent of CGs dominated by red galaxies. On the other hand, less than 50 per cent of CGs are dominated by early-type galaxies, regardless of the environment. The smallest percentage of CGs dominated by early-type galaxies is seen in Non-Embedded CGs (24 per cent). 
The stronger signs of suppression in the star formation of galaxies (redder colours) in CGs that inhabit highly dense structures might be a consequence of the different CGs physical properties previously reported as a function of the environment.  


Finally, we studied whether the galaxy members are affected by the local and global environment. We analysed the fraction of red and early-type galaxies as a function of the absolute magnitude of the galaxies in the r-band. 
This study was performed for each environment to compare the fraction of red/early-type galaxies in CGs with that in the corresponding environment. 
We observed that the fractions of red galaxies are, in general, higher for CGs than for the other structures or the Field (Non-Embedded galaxies) in most of the range of absolute magnitudes, and those differences tend to increase towards less dense environments. 
The variation of these fractions as a function of galaxy luminosity is more notorious in low-density environments than in their high-density counterparts. 
Comparing only galaxies in CGs, there is a tendency for CGs inhabiting denser environments to display a larger fraction of red galaxies as a function of the galaxy luminosity. 
When analysing the fraction of early-type galaxies, similar differences are observed when comparing between CGs and the corresponding environment: CGs show higher fractions of early-type galaxies, but in this case, the decreasing behaviour of the fraction as a function of absolute magnitude is considerably steep in all environments. 
However, when comparing the fraction of early-type galaxies only in CGs in different environments we do not observe distinguishable differences.

These results add another piece to the puzzle that describes the evolution of galaxies in CGs: the importance of the inner local environment defined by the CGs themselves. The main differences observed when comparing the fractions of red and early-type galaxies in CGs with those inhabiting the corresponding environment are due to processes that occur inside CGs. The influence of the global environment, i.e., of the place where the CGs are located, can be seen mainly in how the global physical properties of CGs change with the environment, which is probably the cause of differences observed in the fraction of red galaxies in CGs when comparing high and low-density environments.

One of the main objectives pursued by Hickson's criteria was to define CGs as isolated structures in the Universe.  
As it has been stated in several previous works, in this work we have reinforced the evidence that almost half of the Hickson like CGs obtained using automatic algorithms are not isolated, and even more, we have specified in what type of structures it is more feasible to find them. 
This result leads us to conclude that if we are interested mainly in the physical processes that happen only because of the particular habitat of CGs, then low-density environments are the ideal laboratories to study the true influence of the inner CG environment. Therefore, all the results obtained in this work related to CGs considered Non-Embedded are closer to achieving the goal of highlighting the consequences of an extreme environment such as CGs on their member galaxies. 
However, it is likely that more dynamic and intriguing evolutionary histories should be expected when looking at CGs embedded in high-density environments of the Universe.

As a natural continuation of this research work, we plan to complement this project using mock catalogues. We will investigate whether the location of CGs in and around cosmological structures could be reproduced by the current galaxy formation models, and whether it is a function of the cosmological model and/or the semi-analytical model used to build the synthetic galaxies.

\section*{Acknowledgements}
This publication uses as a parent catalogue the SDSS Data Release 16 (DR16) which is one of the latest data releases of the SDSS-IV.
Funding for the Sloan Digital Sky 
Survey IV has been provided by the 
Alfred P. Sloan Foundation, the U.S. 
Department of Energy Office of 
Science, and the Participating 
Institutions. SDSS-IV acknowledges support and resources from the Center for High 
Performance Computing  at the 
University of Utah. The SDSS 
website is \url{www.sdss.org}.
SDSS-IV is managed by the 
Astrophysical Research Consortium 
for the Participating Institutions 
of the SDSS Collaboration including 
the Brazilian Participation Group, 
the Carnegie Institution for Science, 
Carnegie Mellon University, Center for 
Astrophysics | Harvard \& 
Smithsonian, the Chilean Participation 
Group, the French Participation Group, 
Instituto de Astrof\'isica de 
Canarias, The Johns Hopkins 
University, Kavli Institute for the 
Physics and Mathematics of the 
Universe (IPMU) / University of 
Tokyo, the Korean Participation Group, 
Lawrence Berkeley National Laboratory, 
Leibniz Institut f\"ur Astrophysik 
Potsdam (AIP),  Max-Planck-Institut 
f\"ur Astronomie (MPIA Heidelberg), 
Max-Planck-Institut f\"ur 
Astrophysik (MPA Garching), 
Max-Planck-Institut f\"ur 
Extraterrestrische Physik (MPE), 
National Astronomical Observatories of 
China, New Mexico State University, 
New York University, University of 
Notre Dame, Observat\'ario 
Nacional / MCTI, The Ohio State 
University, Pennsylvania State 
University, Shanghai 
Astronomical Observatory, United 
Kingdom Participation Group, 
Universidad Nacional Aut\'onoma 
de M\'exico, University of Arizona, 
University of Colorado Boulder, 
University of Oxford, University of 
Portsmouth, University of Utah, 
University of Virginia, University 
of Washington, University of 
Wisconsin, Vanderbilt University, 
and Yale University.

This work has been partially supported by Consejo Nacional de Investigaciones Cient\'\i ficas y T\'ecnicas de la Rep\'ublica Argentina (CONICET) and the Secretar\'\i a de Ciencia y Tecnolog\'\i a de la Universidad de C\'ordoba (SeCyT).
\section*{Data Availability} 
The main galaxy catalogue of the SDSS DR16 was downloaded from \url{https://skyserver.sdss.org/casjobs/}.
The sample to minimize the redshift incompleteness, mainly for bright galaxies, is a compiled sample of SDSS DR12 downloaded from \url{http://cosmodb.to.ee/}.

The compact group data used in this work are publicly available at 
\url{https://cdsarc.cds.unistra.fr/viz-bin/cat/J/MNRAS/514/1231}.
The derived data generated in this research will be shared at reasonable request with the corresponding authors.


\bibliographystyle{mnras}
\bibliography{mnras} 

\appendix

\section{Classification of galaxies}

\subsection{Determination of red galaxy threshold}
\label{ap:pasivas}
In this section, we detail the analysis performed for the selection of red galaxies from the colour-magnitude diagram ($M_r$ vs $u-r$).

Due to the colour bi-modality of galaxies, they are often divided into red and blue, using a unique threshold in the colour distribution. To perform a better selection of red galaxies, we consider the absolute magnitudes. Following the methodology used by \cite{zandivarez11}, we divide the whole range of r-band absolute magnitudes in nine bins, and we study the colour $u-r$ distribution of galaxies within each bin. 

\begin{figure}
\begin{center}
\centering
\includegraphics[width=0.98\columnwidth]{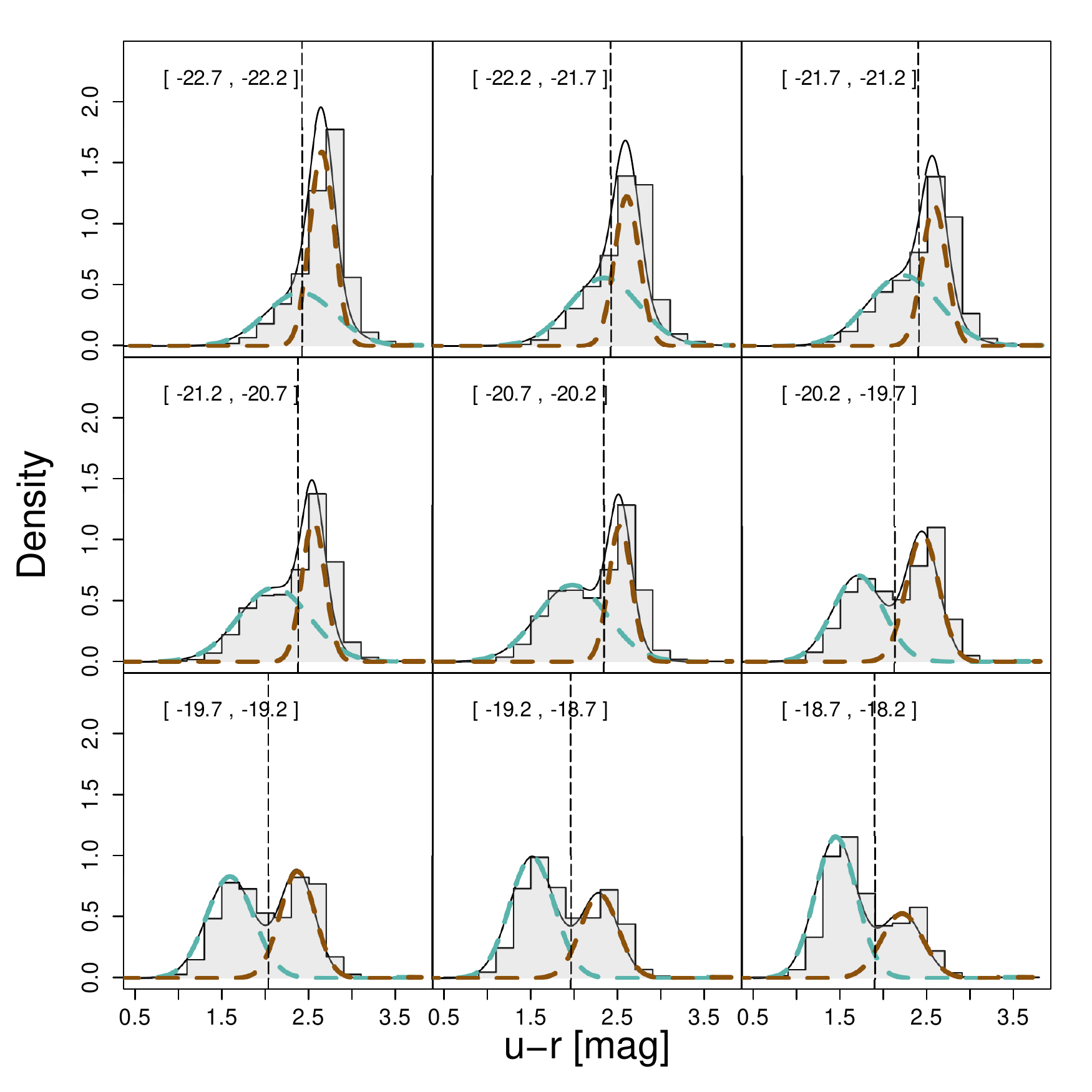}
\includegraphics[width=0.98\columnwidth]{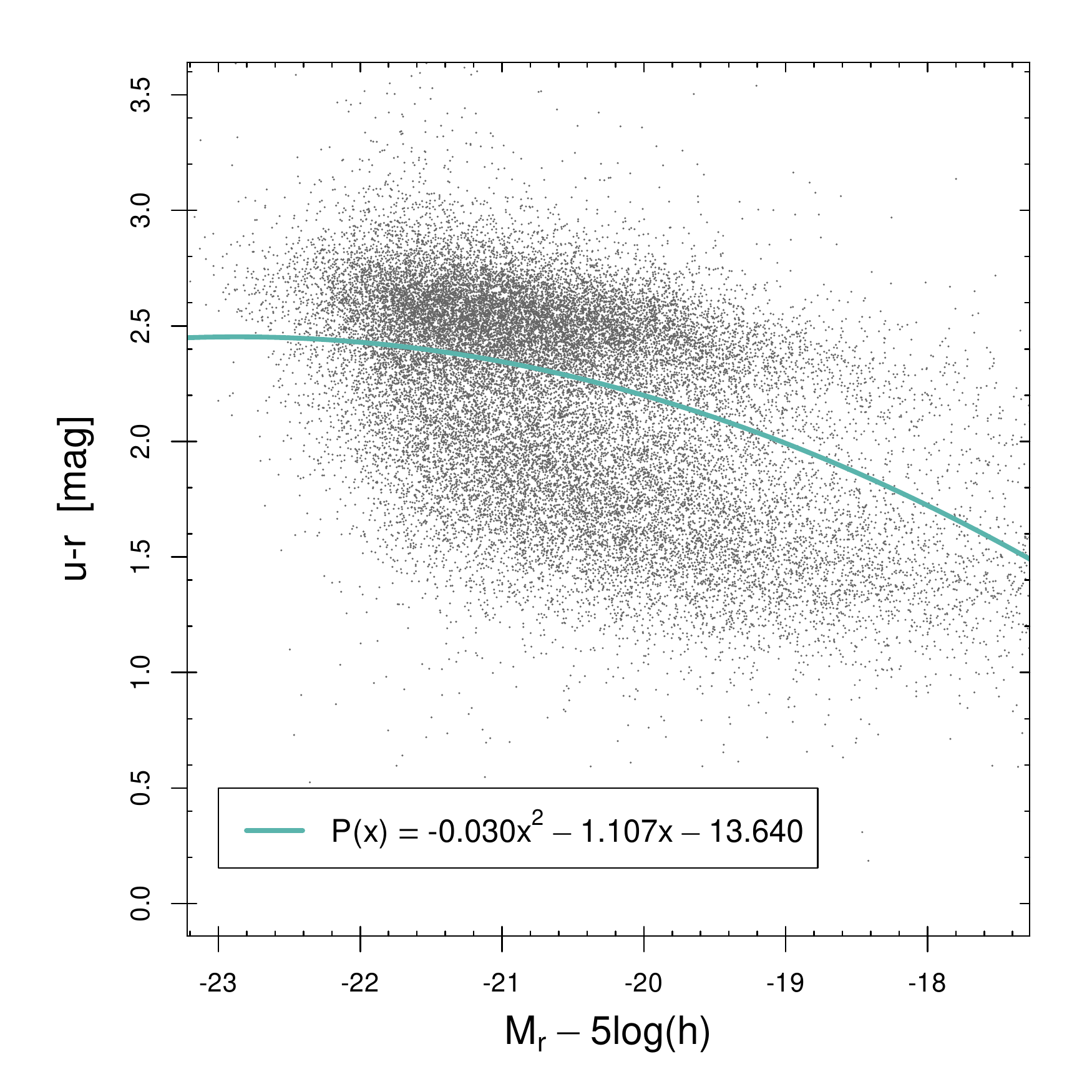}
\caption{{\it Upper panel:} Histograms of the bimodal colour distribution for galaxies split in nine bins of absolute magnitude. Brown and cyan dashed lines represent the Gaussian fits. Vertical dashed lines indicate the intersection of both Gaussians. {\it Bottom panel:} Colour-magnitude diagram for galaxies in SDSS DR16+. The cyan solid line is the function we use to split galaxies into red and blue subsamples.
}
\label{fig:passive_cut}
\end{center}
\end{figure}

We used a method of density estimation based on parameterised finite Gaussian mixture models provided by the {\it Mclust package} of R software \citep{mclust+16}. We select 2 mixture components (gausssians) to perform the uni-variate estimation.
In the top panel of Fig.~\ref{fig:passive_cut}, we show the distributions of galaxy colours for each bin of absolute magnitudes, and we show the Gaussian fits in dashed lines. 
We select the colour value of the intersection of these two Gaussian functions and the middle of the bin of absolute magnitude to characterise each bin and perform a curve fitting.
In the bottom panel of Fig.~\ref{fig:passive_cut}, we show the colour-magnitude diagram of all galaxies. To fit we use Fitting Linear Model $lm$ provided by {\it stats package} \citep{stats+20} and use a 2-degree polynomial to describe the relationship of our data.
Then, the  best fit is the quadratic function: $P(x)= -0.03077 x^2 - 1.4074 x - 13.64045$ with $x=M_r-5\log(h)$ (solid line). We consider that a galaxy belongs to the red galaxy population if its $u-r$ colour is greater than $P(x)$ for the corresponding absolute magnitude.


\subsection{Selection of early-type and late-type galaxies}
\label{ap:early-type}

To split galaxy samples into early and late types, we use the concentration index $C$. This parameter is the ratio between the Petrosian radii enclosing 90 and 50 per cent of the Petrosian flux $C=r90/r50$ and allows us to determine if a galaxy is dominated by the bulge luminosity. Galaxies with $C>2.6$ are defined as early-type, otherwise as late-type \citep{Strateva01}.

\begin{figure}
\begin{center}
\centering
\includegraphics[width=0.98\columnwidth]{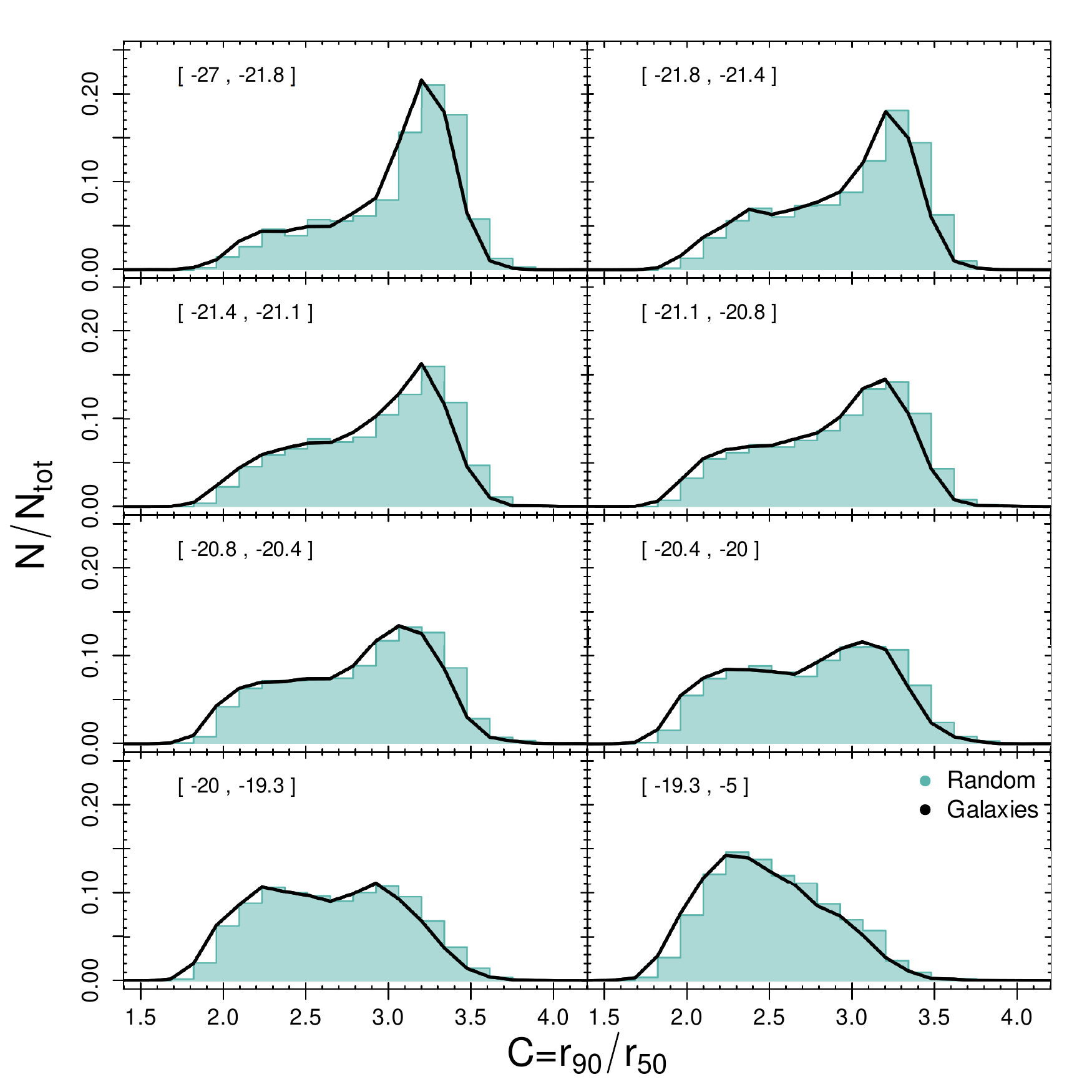}
\caption{Concentration parameter distributions of galaxies with reliable measurement of $r_{50}$ and r-band apparent magnitude $r<16$ (black curve), and for galaxies with randomly assigned C parameter (cyan histogram), for each bin of absolute magnitude.
}
\label{fig:early-type}
\end{center}
\end{figure}

Due to the average seeing in the SDSS being $1.5''$, there are many galaxies whose $r50$ are lower than this value, and therefore unreliable ($\sim 12$ per cent). To deal with this problem, we assign random values to the galaxies with $r_{50}<1.5''$ by following the procedure described by \cite{zandivarez11}. We build a sample of galaxies with reliable Petrosian radii, i.e, $r50>1.5''$ and r-band apparent magnitudes $r<16$, and from this sample, we calculate the cumulative distribution function of the concentration parameter for different bins of absolute magnitude. Using this cumulative function, for each bin of absolute magnitude, every galaxy with unreliable $r_{50}$ in that bin was randomly assigned a value of $"C"$. We repeat this procedure 30 times, and we are left with the distribution that best fits the sample using the Kolmogorov-Smirnov test. In Fig.~\ref{fig:early-type} we show the distribution of C-parameters per bin of absolute magnitude, for both, the reliable measurements (dark curve) and the randomly assigned best values for galaxies without a reliable estimate (cyan histogram).

\appendix


\bsp	
\label{lastpage}
\end{document}